\magnification=1100

\hsize 17truecm
\vsize 23truecm

\font\twelvec=msbm10 at 12pt
\font\sevenc=msbm10 at 9pt
\font\fivec=msbm10 at 7pt

\newfam\co
\textfont\co=\twelvec
\scriptfont\co=\sevenc
\scriptscriptfont\co=\fivec

\def\char{\mathop{\rm Char}\nolimits}
\def\Const{\mathop{\rm Const.}\nolimits}
\def\const{\mathop{\rm const.}\nolimits}

\def\det{\mathop{\rm det}\nolimits}

\def\exp{\mathop{\rm exp}\nolimits}

\def\FS{\mathop{\rm FS}\nolimits}

\def\Id{\mathop{\rm Id}\nolimits}
\def\im{\mathop{\rm Im}\nolimits}

\def\lim{\mathop{\rm lim}\nolimits}

\def\re{\mathop{\rm Re}\nolimits}
\def\supp{\mathop{\rm supp}\nolimits}
\def\Sum{\displaystyle\sum}

\def\sgn{\mathop{\rm sgn}\nolimits}

\def\WF{\mathop{\rm WF}\nolimits}
\def\e{\mathop{\rm \varepsilon}\nolimits}

\baselineskip 15pt


\centerline{\bf BOHR-SOMMERFELD QUANTIZATION RULES REVISITED:} 

\centerline{\bf THE METHOD OF POSITIVE COMMUTATORS}
\bigskip
\centerline{Abdelwaheb IFA ${}^{1}$, Hanen LOUATI ${}^{1,2}$ {\it\&} Michel ROULEUX ${}^{2}$}
\bigskip             
\centerline {${}^{1}$ Universit\'e de Tunis El-Manar, D\'epartement de Math\'ematiques, 1091 Tunis, Tunisia}

\centerline {e-mail: louatihanen42@yahoo.fr, abdelwaheb.ifa@fsm.rnu.tn}

\centerline {${}^{2}$ Aix Marseille Univ, Univ Toulon, CNRS, CPT, Marseille, France} 

\centerline {e-mail: rouleux@univ-tln.fr}  
\bigskip
\noindent {\bf Abstract}:
We revisit the well known Bohr-Sommerfeld quantization rule (BS) of order 2 for a self-adjoint 1-D $h$-Pseudo-differential 
operator within the algebraic and microlocal framework of Helffer and Sj\"ostrand;
BS holds precisely when
the Gram matrix consisting of scalar products of some WKB solutions with respect to the ``flux norm'' is not invertible.
The interest of this procedure lies in its possible generalization to matrix-valued Hamiltonians, like Bogoliubov-de Gennes Hamiltonian.
It is simplified in the scalar case by using action-angle variables. 
\bigskip
\noindent {\bf 0. Introduction}.
\smallskip
Let $p(x,\xi;h)$ be a smooth real classical Hamiltonian on $T^*{\bf R}$~; we will assume that
$p$ belongs to the space of symbols $S^0(m)$ for some order function $m$ with
$$S^N(m)=\{p\in C^\infty(T^*{\bf R}): \forall\alpha\in{\bf N}^{2}, \exists C_\alpha>0,
|\partial^\alpha_{(x,\xi)}p(x,\xi;h)|\leq C_\alpha h^N m(x,\xi)\}\leqno(0.1)$$
and has the semi-classical expansion
$$p(x,\xi; h) \sim p_0(x,\xi)+hp_1(x,\xi)+\cdots, h\rightarrow 0\leqno(0.2)$$
We call as usual $p_0$ the principal symbol, and $p_1$ the sub-principal symbol. 
We also assume that $p+i$ is elliptic. 
This allows to take Weyl quantization of $p$
$$P(x,hD_x;h)u(x;h)=p^w(x,hD_x;h)u(x;h)=(2\pi h)^{-1}\int\int e^{i(x-y)\eta/h} p({x+y\over2},\eta;h)u(y)\,dy\,d\eta\leqno(0.3)$$
so that $P(x,hD_x;h)$ is essentially self-adjoint on $L^2({\bf R})$. 
In case of Schr\"odinger operator $P(x,hD_x)=(hD_x)^2+V(x)$,
$p(x,\xi;h)=p_0(x,\xi)=\xi^2+V(x)$. 
We make the geometrical hypothesis of [CdV1], namely:

Fix some compact interval $I = [E_-,E_+], E_- < E_+$, and assume
that there exists a topological ring ${\cal A}\subset p_0^{-1}(I)$ such that $\partial{\cal A} = {\cal A}_-\cup {\cal A}_+$ 
with ${\cal A}_\pm$ a
connected component of $p_0^{-1}(E_\pm)$. 
Assume also that $p_0$ has no critical point in ${\cal A}$, and
${\cal A}_-$ is included in the disk bounded by ${\cal A}_+$ (if it is not the
case, we can always change $p$ to $-p$.) That hypothesis will be referred in the sequel as Hypothesis (H).

We define the microlocal well $W$ as the disk bounded by ${\cal A}_+$. 
For $E\in I$, let $\gamma_E\subset W$ be a periodic orbit in the energy surface $\{p_0(x,\xi)=E\}$, so that $\gamma_E$ is an
embedded Lagrangian manifold. 

Then if $E_+<E_0=\liminf_{|x,\xi|\to\infty}p_0(x,\xi)$, all eigenvalues of $P$ in $I$ 
are indeed given by {\it Bohr-Sommerfeld quantization condition} (BS) that we recall here, when computed at second order:
\medskip
\noindent {\bf Theorem 0.1}: {\it With the notations and hypotheses stated above, for $h>0$ small enough
there exists a smooth function 
${\cal S}_{h}: I\to {\bf R}$, called the semi-classical action, 
with asymptotic expansion ${\cal S}_{h}(E)\sim S_0(E)+hS_1(E)+h^2S_2(E)+\cdots$
such that $E\in I$ is an eigenvalue of $P$ iff it satisfies the implicit equation 
(Bohr-Sommerfeld quantization condition)
${\cal S}_{h}(E)=2\pi nh$, $n\in{\bf Z}$. The semi-classical action consists of~:

\noindent (i) the classical action along $\gamma_E$
$$S_0(E)=\oint_{\gamma_E}\xi(x)\,dx=\int\int_{\{p_0\leq E\}\cap W}\,d\xi\wedge\,dx$$
(ii) Maslov correction and the integral of the sub-principal 1-form $p_1\,dt$
$$S_1(E)=\pi-\int p_1(x(t),\xi(t))|_{\gamma_E}\,dt$$
(iii) the second order term
$$S_2(E)={1\over24}{d\over dE}\int_{\gamma_E}\Delta\, dt-
\int_{\gamma_E}p_2\,dt-{1\over2}{d\over dE}\int_{\gamma_E}p_1^2\, dt$$
where}
$$\Delta(x,\xi)={\partial^{2} p_{0}\over\partial x^{2}}{\partial^{2} p_{0}\over\partial \xi^{2}}-
\bigl({\partial^{2} p_{0}\over\partial x\,\partial \xi}\bigr)^{2}$$
\smallskip
We recall that $S_3(E)=0$. In constrast with the convention of [CdV], our integrals are oriented integrals, $t$ denoting the
variable in Hamilton's equations. This explains why, in our expressions for $S_2(E)$, derivatives with respect to $E$
(the conjugate variable to $t$) of such integrals have the opposite sign to the corresponding ones in [CdV]. 
See also [IfaM'haRo].   

There are lots of ways to derive BS: the method of matching of WKB solutions [BenOrz], known also as Liouville-Green method [Ol],
which has received many improvements, see e.g. [Ya]; 
the method of the monodromy operator, see [HeRo] and references therein; the method of quantization deformation
based on Functional Calculus and Trace Formulas [Li], [CdV1], [CaGra-SazLiReiRios], [Gra-Saz], [Arg]. 
Note that the method of quantization deformation already assumes BS, it gives only a very convenient way to derive it.
In the real analytic case, 
BS rule, and also tunneling expansions, can be obtained using the so-called ``exact WKB method'' see e.g. [Fe], [DePh], [DeDiPh]
when $P$ is Schr\"odinger operator.  

Here we present still another derivation of BS,
based on the construction of a Hermitian vector bundle of quasi-modes as in [Sj2], [HeSj]. 
Let $K_h^N(E)$ be the microlocal kernel of $P-E$ of order $N$,
i.e. the space of microlocal solutions of $(P-E)u={\cal O}(h^{N+1})$ along the covering of $\gamma_E$ (see Appendix for 
a precise definition). 
The problem is to 
find the set of $E=E(h)$ such that $K_h^N(E)$ contains a global section, i.e. to construct 
a sequence of quasi-modes (QM) $(u_n(h),E_n(h))$ 
of a given order $N$ (practically $N=2$). 
As usual we denote by $K_h(E)$ the microlocal kernel of $P-E$ mod ${\cal O}(h^\infty)$~; since
the distinction between $K_h^N(E)$ and $K_h(E)$ plays no important r\^ole here, we shall be content to write $K_h(E)$.

Actually the method of [Sj2], [HeSj] was elaborated in case of a separatrix, and extends easily
to mode crossing in Born-Oppenheimer type Hamiltonians 
as in [B], [Ro], but somewhat surprisingly it turns out to be harder to set up
in case of a regular orbit, due to ''translation invariance'' of the Hamiltonian flow. In the present scalar 
case, when carried to second order, our method is also more intricated than [Li], [CdV1] and its 
refinements [Gra-Saz] for higher order $N$; nevertheless it shows most useful for matrix valued operators with double characteristics
such as Bogoliubov-de Gennes Hamiltonian [DuGy] (see [BenIfaRo], [BenMhaRo]). 
This method also extends to the scalar case in higher dimensions for a periodic orbit (see [SjZw], [FaLoRo], [LoRo]). 

The paper is organized as follows: 

In Sect.1 we present the main idea of the argument on a simple example, and
recall from [HeSj], [Sj2] the definition of the {\it microlocal Wronskian}.

In Sect.2 we compute BS at lowest order in the special case of Schr\"odinger operator by means of 
microlocal Wronskian and Gram matrix.

In Sect.3 we proceed to more general constructions in the case of 
$h$-Pseudodifferential operator (0.1) so to recover BS at order 2. 

In Sect.4 we use a simpler formalism based on action-angle variables, but which would not
extend to systems such as Bogoliubov-de Gennes Hamiltonian. 

In Sect.5, following [SjZw], we recall briefly the well-posedness of Grushin problem, which shows in particular that there is no other
spectrum in $I$ than this given by BS.

At last, the Appendix accounts for a short introduction to microlocal and semi-classical Analysis used in the main text. 
\bigskip
\noindent {\it Acknowledgements}: We thank a referee for his constructive remarks. This work has been partially supported by
the grant PRC CNRS/RFBR 2017-2019 No.1556 ``Multi-dimensional semi-classical problems of Condensed Matter Physics and Quantum Dynamics''.


\bigskip
\noindent {\bf 1. Main strategy of the proof}.
\smallskip
The best algebraic and microlocal framework for computing quantization rules in the self-adjoint case,
cast in the fundamental works [Sj2], [HeSj], is based on Fredholm theory,
and the classical ``positive commutator method'' using conservation of some quantity 
called a ``quantum flux''. 
\medskip
\noindent {\it a) A simple example}
\smallskip
As a first warm-up, consider $P=hD_x$ acting on $L^2({\bf S}^1)$ with periodic boundary condition
$u(x)=u(x+2\pi)$. It is well-known that $P$ has discrete spectrum $E_k(h)=kh$, $k\in{\bf Z}$, with eigenfunctions
$u_k(x)=(2\pi)^{-1/2}e^{ikx}=(2\pi)^{-1/2}e^{iE_k(h)x/h}$. Thus BS quantization rule can be written as $\oint_{\gamma_E}\xi\,dx=2\pi kh$, where 
$\gamma_E=\{x\in {\bf S}^1; \xi=E\}$. 

We are going to derive this result using the monodromy properties of the solutions of $(hD_x-E)u=0$.
For notational convenience, we change energy variable $E$ into $z$. 
Solving for $(P-z)u(x)=0$, we get two solutions with the same expression but defined on different charts
$$u^a(x)=e^{izx/h}, -\pi<x<\pi, \quad u^{a'}(x)=e^{izx/h}, 0<x<2\pi\leqno(1.1)$$
indexed by angles $a=0$ and $a'=\pi$ on ${\bf S}^1$. In the following we take advantage of the fact that 
these functions differ but when $z$ belongs to the spectrum of $P$. 

Let also $\chi^a\in C_0^\infty({\bf S}^1)$
be equal to 1 near $a$, $\chi^{a'}=1-\chi^a$. We set $F^a_\pm={i\over h}[P,\chi^a]_\pm u^a$, where $\pm$ denotes the part of the 
commutator supported in the half circles $0<x<\pi$ and $-\pi<x<0$ mod $2\pi$. Similarly  
$F^{a'}_\pm={i\over h}[P,\chi^{a'}]_\pm u^{a'}$.
We compute
$$(u^a|F^a_+)=\bigl(u^a|(\chi^a)'u^a\bigr)=\int_0^\pi(\chi^a)'(x)\,dx=\chi^a(\pi)-\chi^a(0)=-1$$
Similarly $(u^a|F^a_-)=1$, and also replacing $a$ by $a'$ so that
$$(u^a|F^a_+-F^a_-)=-2, \quad (u^{a'}|F^{a'}_+-F^{a'}_-)=2\leqno(1.2)$$
We evaluate next the crossed terms $(u^{a'}|F^a_+-F^a_-)$ and $(u^a|F^{a'}_+-F^{a'}_-)$. 
Since $u^{a'}(x)=u^a(x)=e^{izx/h}$ on the upper-half circle (once embedded into the complex plane), 
and $u^a(x)=e^{izx/h}, u^{a'}(x)=e^{iz(x+2\pi)/h}$ on the lower-half circle we have
$$(u^{a'}|F^a_+-F^a_-)=\int_0^\pi e^{izx/h}(\chi^a)'e^{-izx/h}\,dx-\int_{-\pi}^0e^{iz(x+2\pi)/h}(\chi^a)'e^{-izx/h}\,dx$$
We argue similarly
for $(u^{a}|F^{a'}_+-F^{a'}_-)$, using also that 
$(\chi^{a'})'=-(\chi^a)'$. So we have
$$(u^{a'}|F^{a}_+-F^{a}_-)=-1-e^{2i\pi z/h}, \quad (u^{a}|F^{a'}_+-F^{a'}_-)=1+e^{-2i\pi z/h}\leqno(1.3)$$
It is convenient to view $F^a_+-F^a_-$ and $F^{a'}_+-F^{a'}_-$ as belonging to co-kernel of $P-z$ in the sense they are not
annihilated by $P-z$. So we form Gram matrix 
$$G^{(a,a')}(z)=\pmatrix{(u^a|F^{a}_+-F^{a}_-)&(u^{a'}|F^{a}_+-F^{a}_-)\cr (u^a|F^{a'}_+-F^{a'}_-)&(u^{a'}|F^{a'}_+-F^{a'}_-)\cr}\leqno(1.4)$$
and an elementary computation using (1.2) and (1.3) shows that 
$$\det G^{(a,a')}(z)=-4\sin^2(\pi z/h)$$
so the condition that $u^a$ coincides with $u^{a'}$ is precisely that $z=kh$, with $k\in{\bf Z}$.

Next we investigate Fredholm properties of $P$ as in [SjZw], recovering the fact that $h{\bf Z}$ is the only spectrum of $P$. 

Notice that (1.4) is not affected when multiplying $u^{a'}$ by a phase factor,
so we can replace $u^{a'}$ by $e^{-iz\pi/h}u^{a'}$. 
Starting from the point $a=0$ we associate with $u^a$ the multiplication operator $v_+\mapsto I^a(z)v_+=u^a(x)v_+$ on ${\bf C}$,
i.e. Poisson operator with  
``Cauchy data'' $u(0)=v_+\in{\bf C}$. Define the ``trace operator''
$R_+(z)u=u(0)$. 

Similarly multiplication by $u^{a'}$ defines Poisson operator $I^{a'}(z)v_+=u^{a'}(x)v_+$, 
which has the same ``Cauchy data'' $v_+$ at $a'=\pi$ as $I^a(z)$ at $a=0$.

Consider the multiplication operators
$$E_+(z)=\chi^a I^{a}(z)+(1-\chi^a)e^{i\pi z/h}I^{a'}(z), \quad R_-(z)={i\over h}[P,\chi^a]_-I^{a'}(z), \quad E_{-+}(z)=2h\sin(\pi z/h)$$
We claim that 
$$(P-z)E_+(z)+R_-(z)E_{-+}(z)=0\leqno(1.5)$$
Namely as before (but after we have replaced 
$u^{a'}$ by $e^{-iz\pi/h}u^{a'}$) evaluating on $0<x<\pi$, 
we have $I^a(z)=e^{ixz/h}, I^{a'}(z)=e^{-i\pi z/h}e^{ixz/h}$, while evaluating on $-\pi<x<0$, 
$I^a(z)=e^{ixz/h}, I^{a'}(z)=e^{-i\pi z/h}e^{i(x+2\pi)z/h}$. Now 
$(P-z)E_+(z)=[P,\chi^a]\bigl(I^a(z)-e^{i\pi z/h}I^{a'}(z)\bigr)$ vanishes on $0<x<\pi$, while is precisely equal to
$2h\sin(\pi z/h){i\over h}[P,\chi^a]I^{a'}(z)$ on $-\pi<x<0$. So (1.5) follows. 

Hence the Grushin problem
$${\cal P}(z;h){u\choose u_-}=\pmatrix{P-z&R_-(z)\cr R_+(z)&0}{u\choose u_-}={v\choose v_+}\leqno(1.6)$$
with $v=0$ has a solution $u=E_+(z)v_+$, $u_-=E_{-+}(z)v_+$, with $E_{-+}(z)$ the effective Hamiltonian. Following [SjZw] one can show
that with this choice of $R_\pm(z)$, problem (1.6) is well posed, ${\cal P}(z)$ is invertible, 
and 
$${\cal P}(z)^{-1}=\pmatrix{E(z)&E_+(z)\cr E_-(z)&E_{-+}(z)}\leqno(1.7)$$ 
with
$$(P-z)^{-1}=E(z)-E_+(z)E_{-+}(z)^{-1}E_-(z)\leqno(1.8)$$
Hence  $z$ is an eigenvalue of $P$ iff $E_{-+}(z)=0$,
which gives the spectrum $z=kh$ as expected.

These Fredholm properties have been further generalized to a periodic orbit in higher dimensions in
several ways [SjZw], [NoSjZw], [FaLoRo] where $E_{-+}(z)$ is defined by means of 
the monodromy operator as $E_{-+}(z)=\Id-M(z)$ (in this example $M(z)=e^{2i\pi z/h}$). In fact our argument here differs essentially from 
the corresponding one in [SjZw]
by the choice of cutt-off $\chi^a$. 
We have considered functions on ${\bf S}^1$, but in Sect.4,
we work on the covering of ${\bf S}^1$ instead, using a single Poisson operator.
\medskip
\noindent {\it b) The microlocal Wronskian}.
\smallskip
We now consider Bohr-Sommerfeld on the real line. Contrary to the periodic case that we have just investigated,
where Maslov index is $m=0$, we get in general $m=2$ for BS on the real line, as is the case for the
harmonic oscillator $P=(hD_x)^2+x^2$ on $L^2({\bf R})$. Otherwise, the argument is pretty much the same. 

Bohr-Sommerfeld quantization rules result in constructing quasi-modes by WKB approximation along a closed Lagrangian manifold
$\Lambda_E\subset\{p_0=E\}$, i.e. a periodic orbit of Hamilton vector field $H_{p_0}$ with energy $E$.
This can be done locally according to the rank of the projection $\Lambda_E\to{\bf R}_x$.

Thus the set $K_h(E)$ of asymptotic solutions to $(P-E)u=0$ along (the covering of)
$\Lambda_E$ can be considered as a bundle over ${\bf R}$
with a compact base, corresponding to the ``classically allowed region'' at energy $E$. 
The sequence of eigenvalues
$E=E_n(h)$ is determined by the condition that the resulting quasi-mode, gluing together asymptotic
solutions from different coordinates patches along $\Lambda_E$,
be single-valued, i.e. $K_h(E)$ have trivial holonomy. 

Assuming $\Lambda_E$ is smoothly embedded in $T^*{\bf R}$, it can always be parametrized by a non degenerate phase function. 
Of particular interest are the critical points of the phase functions, or {\it focal points}
which are responsible for the change in Maslov index. Recall that $a(E)=(x_E,\xi_E)\in\Lambda_E$ 
is called a focal point if $\Lambda_E$ ``turns vertical'' at $a(E)$, i.e.
$T_{a(E)}\Lambda_E$ is no longer transverse to the fibers $x=\Const $ in $T^*{\bf R}$. 
In any case, however, $\Lambda_E$ can be parametrized locally either by a phase $S=S(x)$ (spatial representation) or a phase 
$\widetilde S=\widetilde S(\xi)$
(Fourier representation). Choose an orientation on $\Lambda_E$ and for $a\in\Lambda_E$ (not necessarily a focal point), denote by
$\rho=\pm1$ its oriented segments near $a$. 
Let $\chi^a\in C_0^\infty({\bf R}^2)$ be a smooth cut-off equal to 1 near $a$, and $\omega^a_\rho$ a small neighborhood
of $\supp [P,\chi^a]\cap\Lambda_E$ near $\rho$.
Here the notation $\chi^a$ holds for $\chi^a(x,hD_x)$ as in (0.3), and 
we shall write $P(x,hD_x)$ (spatial representation) as well as $P(-hD_\xi,\xi)$ (Fourier representation).
Recall that unitary $h$-Fourier transform for a semi-classical distribution $u(x;h)$ 
is given by $\widehat u(\xi;h)=(2\pi h)^{-1/2}\int e^{-ix\xi/h}u(x;h)\,dx$ (see Appendix for a review of semi-classical asymptotics).
\medskip
\noindent{\bf Definition 1.1}: {\it Let $P$ be self-adjoint, and $u^a,v^a\in K_h(E)$ be supported microlocally on $\Lambda_E$. We call 
$${\cal W}^a_\rho(u^a,\overline{v^a})=\bigl({i\over h}[P,\chi^a]_\rho u^a|v^a\bigr)\leqno(1.10)$$ 
the {\it microlocal Wronskian} of $(u^a,\overline{v^a})$ in $\omega^a_\rho$. Here ${i\over h}[P,\chi^a]_\rho$ denotes the part of the 
commutator supported on $\omega^a_\rho$.  }
\smallskip
To understand that terminology, let $P=-h^2\Delta+V$, $x_E=0$ and change $\chi$ to Heaviside unit step-function $\chi(x)$,
depending on $x$ alone. Then in distributional sense,
we have ${i\over h}[P,\chi]=-ih\delta'+2\delta hD_x$, where $\delta$ denotes the Dirac measure at 0, and $\delta'$ its derivative,
so that $\bigl({i\over h}[P,\chi]u|u\bigr)=-ih\bigl(u'(0)\overline{u(0)}-u(0)\overline{u'(0)}\bigr)$
is the usual Wronskian of $(u,\overline u)$.
\smallskip
\noindent{\bf Proposition 1.2}: {\it Let $u^a,v^a\in K_h(E)$ be as above, 
and denote by $\widehat u$ the $h$-Fourier (unitary) transform of $u$. Then
$$\leqalignno{
&\bigl({i\over h}[P,\chi^a] u^a|v^a\bigr)=\bigl({i\over h}[P,\chi^a] \widehat u^a|\widehat v^a\bigr)=0&(1.11)\cr
&\bigl({i\over h}[P,\chi^a]_+ u^a|v^a\bigr)=-\bigl({i\over h}[P,\chi^a]_- u^a|v^a\bigr)&(1.12)\cr
}$$ 
all equalities being understood mod ${\cal O}(h^\infty)$, (resp. ${\cal O}(h^{N+1}))$ when considering 
$u^a,v^a\in K_h^N(E)$ instead. Moreover, ${\cal W}^a_\rho(u^a,\overline{v^a})$ does not depend mod ${\cal O}(h^\infty)$ 
(resp. ${\cal O}(h^{N+1})$) on the choice of $\chi^a$ 
as above.}
\smallskip
\noindent {\it Proof}: Since $u^a,v^a\in K_h(E)$ are distributions in $L^2$, the equality
(1.11) follows from Plancherel formula and the regularity of microlocal solutions in $L^2$, 
$p+i$ being elliptic. If $a$ is not a focal point, $u^a,v^a$ are smooth WKB solutions near $a$, so we can
expand the commutator in $w=\bigl({i\over h}[P,\chi^a] u^a|v^a\bigr)$ and use that $P$ is self-adjoint to show that
$w={\cal O}(h^\infty)$. If $a$ is a focal point, $u^a,v^a$ are smooth WKB solutions in Fourier representation, so again
$w={\cal O}(h^\infty)$. Then (1.12) follows
from Definition 1.1. $\clubsuit$
\smallskip
We can find a linear combination of 
${\cal W}^a_\pm$, (depending on $a$) which defines a sesquilinear form on $K_h(E)$, so that this Hermitean form makes 
$K_h(E)$ a metric bundle, endowed with the gauge group $U(1)$. 
This linear combinaison is prescribed as the construction of Maslov index~: namely 
we take ${\cal W}^a(u^a,\overline{u^a})={\cal W}^a_+(u^a,\overline{u^a})-{\cal W}^a_-(u^a,\overline{u^a})>0$ when the critical 
point $a$ of $\pi_{\Lambda_E}$ is traversed in the $-\xi$ direction to the right of the fiber 
(or equivalently ${\cal W}^a(u^a,\overline{u^a})=-{\cal W}^a_+(u^a,\overline{u^a})+{\cal W}^a_-(u^a,\overline{u^a})>0$ 
while traversing $a$ in the $+\xi$ direction to the left of the fiber). Otherwise, just exchange the signs. 
When $\Lambda_E$ is a convex curve, there are only 2 focal points. In general there may be many focal points $a$, 
but each jump of Maslov index is compensated at the next focal point
while traversing to the other side of the fiber (Maslov index is computed mod 4), see [BaWe,Example 4.13]. 

As before our method consists in constructing Gram matrix of a generating system of ${K}_h(E)$ in a suitable dual basis; 
its determinant vanishes precisely at the eigenvalues $E=E_n(h)$. 

Note that when energy surface $p_0=E$ is singular, and $\Lambda_E$ is a separatrix (''figure eight'', or homoclinic case), 
equality (1.12) does not hold
near the ``branching point'', see [Sj2] and its generalization to multi-dimensional case [BoFuRaZe]. 
\bigskip
\noindent{\bf 2. Bohr-Sommerfeld quantization rules in the case of a Schr\"odinger operator}.
\smallskip
As a second warm-up, we derive the well known BS quantization rule using microlocal Wronskians
in case of a potential well, i.e. $\Lambda_E$ has only 2 focal points.
Consider the spectrum of Schr\"odinger operator 
$P(x,hD_x)=(hD_x)^2+V(x)$ near the energy level $E_0<\liminf_{|x|\to\infty}V(x)$, when $\{V\leq E\}=[x'_E,x_E]$
and $x'_E,x_E$ are simple turning points, $V(x'_E)=V(x_E)=E$, $V'(x'_E)<0,V'(x_E)>0$. 
For a survey of WKB theory, see e.g. [Dui], [BaWe] or [CdV]. It is convenient to start the construction from the focal points $a$ or $a'$.
We identify a focal point $a=a_E=(x_E,0)$ 
with its projection $x_E$. 
We know that microlocal solutions $u$ of $(P-E)u=0$ in a (punctured) neighborhood of $a$ are of the form
$$u^a(x,h)={C\over\sqrt2}\bigl(e^{i\pi/4}(E-V)^{-1/4}e^{iS(a,x)/h}+e^{-i\pi/4}(E-V)^{-1/4}e^{-iS(a,x)/h}+{\cal O}(h)\bigr), \ C\in{\bf C}\leqno(2.1)$$
where $S(y,x)=\int_y^x\xi_+(t)\, dt$, and $\xi_+(t)$ is the positive root of $\xi^2+V(t)=E$. 
In the same way, the microlocal solutions of  $(P-E)u=0$ in a (punctured) neighborhood of $a'$ have the form
$$u^{a'}(x,h)={C'\over\sqrt2}\bigl(e^{-i\pi/4}(E-V)^{-1/4}e^{iS(a',x)/h}+e^{i\pi/4}(E-V)^{-1/4}e^{-iS(a',x)/h}+{\cal O}(h)\bigr), \ C'\in{\bf C} 
\leqno(2.2)$$
These expressions result in computing by the method of stationary phase
the oscillatory integral that gives the solution of $(P(-hD_\xi,\xi)-E)\widehat u=0$
in Fourier representation. The change of phase factor $e^{\pm i\pi/4}$ accounts for Maslov index.
For later purposes, we recall here from [H\"o,Thm 7.7.5] that if $f:{\bf R}^d\to{\bf C}$, with $\im f\geq0$ has a non-degenerate
critical point at $x_0$, then 
$$\int_{{\bf R}^{d}}e^{{i\over h}f(x)}\,u(x)\,dx \sim e^{{i\over h}f(x_{0})}\,\bigl(\det({f''(x_{0})\over 2i\pi h})\bigr)^{-1/2}\,
\sum_{j}h^{j}\,L_{j}(u)(x_{0})\leqno(2.3)$$
where $L_{j}$ are linear forms, $L_{0}u(x_{0})=u(x_{0})$, and
$$L_{1}u(x_{0})=\sum_{n=0}^{2}{2^{-(n+1)}\over in!(n+1)!}\langle (f''(x_{0}))^{-1}D_{x},
D_{x}\rangle^{n+1}\bigl((\Phi_{x_{0}})^{n}u)(x_{0})\leqno(2.4)$$
where $\Phi_{x_0}(x)=f(x)-f(x_{0})-{1\over2}\langle f''(x_{0})(x-x_{0}), x-x_{0}\rangle$
vanishes of order 3 at $x_{0}$.
 
For the sake of simplicity, we omit henceforth ${\cal O}(h)$ terms, but the computations below
extend to all order in $h$ (practically, at least for $N=2$), thus giving the asymptotics of BS. This will be elaborated in Section 3.

The semi-classical distributions $u^a,u^{a'}$ span the microlocal kernel $K_h$ of $P-E$ in $(x,\xi)\in]a',a[\times{\bf R}$~; they are 
normalized using microlocal Wronskians as follows.

Let $\chi^a\in C_0^\infty({\bf R}^2)$ as in the Introduction be a smooth cut-off equal to 1 near $a$.
Without loss of generality, we can take $\chi^a(x,\xi)=\chi^a_1(x)\chi_2(\xi)$, so that $\chi_2\equiv1$ on small neighborhoods $\omega^a_\pm$,
of $\supp [P,\chi^a]\cap\{\xi^2+V=E\}$ in $\pm\xi>0$. We define
$\chi^{a'}$ similarly.
Since ${i\over h}[P,\chi^a]=2(\chi^a)'(x)hD_x-ih(\chi^a)''$, by (2.1) and (2.2) we have, mod ${\cal O}(h)$:
$$\eqalign{
&{i\over h}[P,\chi^a]u^a(x,h)=
\sqrt2C(\chi^a_1)'(x)\bigl(e^{i\pi/4}(E-V)^{1/4}e^{iS(a,x)/h}-e^{-i\pi/4}(E-V)^{1/4}e^{-iS(a,x)/h}\bigr)\cr
&{i\over h}[P,\chi^{a'}]u^{a'}(x,h)=
\sqrt2C'(\chi^{a'}_1)'(x)\bigl(e^{-i\pi/4}(E-V)^{1/4}e^{iS(a',x)/h}-e^{i\pi/4}(E-V)^{1/4}e^{-iS(a',x)/h}\bigr)\cr
}$$
Let 
$$F^a_\pm(x,h)={i\over h}[P,\chi^a]_\pm u^a(x,h)=\pm\sqrt2C(\chi^a_1)'(x)e^{\pm i\pi/4}(E-V)^{1/4}e^{\pm iS(a,x)/h}\leqno(2.5)$$ 
so that:
$$\eqalign{
&(u^a|F^a_+-F^a_-)=|C|^2\bigl(e^{i\pi/4}(E-V)^{-1/4}e^{iS(a,x)/h}|(\chi^a_1)'e^{i\pi/4}(E-V)^{1/4}e^{iS(a,x)/h})\cr
&+|C|^2(e^{-i\pi/4}(E-V)^{-1/4}e^{-iS(a,x)/h}|(\chi^a_1)'e^{-i\pi/4}(E-V)^{1/4}e^{-iS(a,x)/h})\bigr)+{\cal O}(h)\cr
&=|C|^2(\int_{-\infty}^a(\chi^a_1)'(x)dx+\int_{-\infty}^a(\chi^a_1)'(x)dx)+{\cal O}(h)=2|C|^2+{\cal O}(h)
}$$
(the mixed terms such as $\bigl(e^{i\pi/4}(E-V)^{-1/4}e^{iS(a,x)/h}|(\chi^a_1)'e^{-i\pi/4}(E-V)^{1/4}e^{-iS(a,x)/h})$ are ${\cal O}(h^\infty)$ because
the phase is non stationary), thus $u^a$ is normalized mod ${\cal O}(h)$
if we choose $C=2^{-1/2}$. 
In the same way, with 
$$F^{a'}_\pm(x,h)={i\over h}[P,\chi^{a'}]_\pm u^{a'}(x,h)=\pm\sqrt2C'(\chi^{a'}_1)'(x)e^{\mp i\pi/4}(E-V)^{1/4}e^{\pm iS(a',x)/h}\leqno(2.6)$$ 
we get
$$(u^{a'}|F^{a'}_+-F^{a'}_-)=|C'|^2(\int_{a'}^\infty(\chi^{a'}_1)'(x)dx+\int_{a'}^\infty(\chi^{a'}_1)'(x)dx)+{\cal O}(h)=-2|C'|^2+{\cal O}(h)$$
and we choose again $C'=C$ which normalizes $u^{a'}$ mod ${\cal O}(h)$. Normalization carries to higher order, as is shown in Sect.3
for a more general Hamiltonian.
 
So there is a natural duality product between $K_h(E)$ and the
span of functions $F^a_+-F^a_-$ and $F^{a'}_+-F^{a'}_-$ in $L^2$. As in [Sj2], [HeSj]
we can show that this space is microlocally transverse to $\im (P-E)$ on $(x,\xi)\in]a',a[\times{\bf R}$, and thus identifies with 
the microlocal co-kernel $K_h^*(E)$ of
$P-E$; in general $\dim K_h(E)=\dim K_h^*(E)=2$, unless $E$ is an eigenvalue, in which case $\dim K_h=\dim K_h^*=1$ (showing that
$P-E$ is of index 0 when Fredholm, which is indeed the case.~)

Microlocal solutions $u^a$ and $u^{a'}$ extend as smooth solutions on the whole interval $]a',a[$; we denote them by $u_1$ and $u_2$.
Since there are no other focal points between $a$ 
and $a'$, they are expressed by the same formulae (which makes the analysis particularly simple) and satisfy~:
$$(u_1|F^{a}_+-F^{a}_-)=1,\quad (u_2|F^{a'}_+-F^{a'}_-)=-1$$
Next we compute (still modulo ${\cal O}(h)$)
$$\eqalign{
&(u_1|F^{a'}_+-F^{a'}_-)={1\over2}(e^{i\pi/4}(E-V)^{-1/4}e^{iS(a,x)/h}|(\chi^{a'}_1)'e^{-i\pi/4}(E-V)^{1/4}e^{iS(a',x)/h})\cr
&+{1\over2}(e^{-i\pi/4}(E-V)^{-1/4}e^{-iS(a,x)/h}|(\chi^{a'}_1)'e^{i\pi/4}(E-V)^{1/4}e^{-iS(a',x)/h})\cr
&={i\over2}e^{-iS(a',a)/h}\int_{a'}^\infty(\chi^{a'}_1)'(x)dx-{i\over2}e^{iS(a',a)/h}\int_{a'}^\infty(\chi^{a'}_1)'(x)dx=-\sin (S(a',a)/h)\cr
}$$
(taking again into account that the mixed terms are ${\cal O}(h^\infty)$). Similarly
$(u_2|F^{a}_+-F^{a}_-)=\sin (S(a',a)/h)$. Now we define Gram matrix
$$G^{(a,a')}(E)=\pmatrix{(u_1|F^{a}_+-F^{a}_-)&(u_2|F^{a}_+-F^{a}_-)\cr (u_1|F^{a'}_+-F^{a'}_-)&(u_2|F^{a'}_+-F^{a'}_-)\cr}\leqno(2.7)$$
whose determinant $-1+\sin^2 (S(a',a)/h)=-\cos^2 (S(a',a)/h)$ vanishes precisely on eigenvalues of $P$ in $I$, so we recover the 
well known BS quantization condition
$$\oint\xi(x)\, dx=2\int_{a'}^a(E-V)^{1/2}\, dx=2\pi h(k+{1\over2})+{\cal O}(h)\leqno(2.8)$$
and $\det G^{(a,a')}(E)$ is nothing but Jost function which is computed e.g. in [DePh], [DeDiPh] by another method. 

\bigskip
\noindent{\bf 3. The general case}
\smallskip
By the discussion after Proposition 1.1, it clearly suffices to consider the case when $\gamma_E$ contains only 2 focal points
which contribute to Maslov index. We shall content throughout to BS mod ${\cal O}(h^2)$.
\medskip
\noindent {\it a) Quasi-modes mod ${\cal O}(h^2)$ in Fourier representation}.
\smallskip
Let $a=a_E=(x_E,\xi_E)$ be such a focal point. Following a well known procedure we can trace back to [Sj1],
we first seek for WKB solutions in Fourier representation near $a$ of the form $\widehat u(\xi)=e^{i\psi(\xi)/h}b(\xi;h)$, see e.g. [CdV2]
and Appendix below. 
Here the phase $\psi=\psi_E$ solves Hamilton-Jacobi equation $p_0(-\psi'(\xi),\xi)=E$, and can be normalized by $\psi(\xi_E)=0$;
the amplitude $b(\xi;h)=b_0(\xi)+hb_1(\xi)+\cdots$ has to be found recursively
together with $a(x,\xi;h)=a_0(x,\xi)+ha_1(x,\xi)+\cdots$, 
such that 
$$hD_\xi\bigl( e^{i(x\xi+\psi(\xi))/h} a(x,\xi;h)\bigr)=P(x,D_x;h)\bigl(e^{i(x\xi+\psi(\xi))/h}b(\xi;h)\bigl)$$
Expanding the RHS by stationary phase (2.3),  we find
$$hD_\xi\bigl( e^{i(x\xi+\psi(\xi))/h} a(x,\xi;h)\bigr)=
e^{i(x\xi+\psi(\xi))/h}b(\xi;h)\bigl(p_0(x,\xi)-E+h\widetilde p_1(x,\xi)+h^2\widetilde p_2(x,\xi)+{\cal O}(h^3)\bigr)$$ 
$p_0$ being the principal symbol of $P$, 
$$\widetilde p_1(x,\xi)=p_1(x,\xi)+{1\over2i}
{\partial^2p_0\over\partial x\partial\xi}(x,\xi), \ 
\widetilde p_2(x,\xi)=p_2(x,\xi)+{1\over2i}
{\partial^2p_1\over\partial x\partial\xi}(x,\xi)-{1\over8}{\partial^4p_0\over\partial x^2\partial\xi^2}(x,\xi)
$$
Collecting the coefficients of ascending powers of $h$, we get
$$\leqalignno{
&(p_0-E)b_0=(x+\psi'(\xi))a_0&(3.1)_0\cr
&(p_0-E)b_1+\widetilde p_1b_0=(x+\psi'(\xi))a_1+{1\over i}{\partial a_0\over\partial\xi}&(3.1)_1\cr
&(p_0-E)b_2+\widetilde p_1b_1+\widetilde p_2b_0=(x+\psi'(\xi))a_2+{1\over i}{\partial a_1\over\partial\xi}&(3.1)_2\cr
}$$
and so on. 
Define $\lambda(x,\xi)$ by $p_0(x,\xi)-E=\lambda(x,\xi)(x+\psi'(\xi))$, we have
$$\lambda(-\psi'(\xi),\xi)=\partial_xp_0(-\psi'(\xi),\xi)=\alpha(\xi)\leqno(3.2)$$
This gives $a_0(x,\xi)=\lambda(x,\xi)b_0(\xi)$ for $(3.1)_0$.
We look for $b_0$ by noticing that $(3.1)_1$ is solvable iff 
$$(\widetilde p_1b_0)|_{x=-\psi'(\xi)}={1\over i}{\partial a_0\over\partial\xi}|_{x=-\psi'(\xi)}$$
which yields the first order ODE $L(\xi,D_\xi)b_0=0$, with
$L(\xi,D_\xi)=\alpha(\xi)D_\xi+{1\over2i}\alpha'(\xi)-p_1(-\psi'(\xi),\xi)$. We find
$$b_0(\xi)=C_0|\alpha(\xi)|^{-1/2}e^{i\int{p_1\over\alpha}}$$
with an arbitrary constant $C_0$. This gives in turn 
$$a_1(x,\xi)=\lambda(x,\xi)b_1(\xi)+\lambda_0(x,\xi)\leqno(3.3)$$ 
with 
$$\lambda_0(x,\xi)={b_0(\xi)\widetilde p_1+i{\partial a_0\over\partial\xi}\over x+\partial_\xi\psi}$$
which is smooth near $a_E$. 
At the next step, we look for $b_1$ by noticing that $(3.1)_2$ is solvable iff 
$$(\widetilde p_1b_1+\widetilde p_2b_0)|_{x=-\psi'(\xi)}={1\over i}{\partial a_1\over\partial\xi}|_{x=-\psi'(\xi)}$$
Differentiating (3.3) gives $L(\xi,D_\xi)b_1=\widetilde p_2b_0+i\partial_\xi\lambda_0|_{x=-\psi'(\xi)}$, which we solve
for $b_1$. We eventually get, mod ${\cal O}(h^2)$
$$\widehat u^a(\xi;h)=(C_0+hC_1+hD_1(\xi))|\alpha(\xi)|^{-1/2}\exp {i\over h}\bigl[\psi(\xi)+h\int_{\xi_E}^\xi
{p_1(-\psi'(\zeta),\zeta)\over\alpha(\zeta)}\,d\zeta]
\leqno(3.4)$$
where we have set (for $\xi$ close enough to $\xi_E$ so that $\alpha(\xi)\neq0$)
$$D_1(\xi)=\sgn(\alpha(\xi_E))\int_{\xi_E}^\xi\exp[-i\int_{\xi_E}^\zeta{p_1\over\alpha}]
\bigl(i\widetilde p_2b_0-\partial_\xi\lambda_0|_{x=-\psi'(\zeta)}\bigr)\, 
|\alpha(\zeta)|^{-1/2}\,d\zeta\leqno(3.5)$$
The integration constants $C_0,C_1$ will be determined 
by normalizing the microlocal Wronskians as follows.
We postpone to Sect.3.c the proof of this Proposition making us of the spatial representation of $u^a$. 
\medskip
\noindent {\bf Proposition 3.1}: {\it With the hypotheses above, the microlocal Wronskian near a focal point $a_E$ is given by}
$$\eqalign{
&{\cal W}^a(u^a,\overline{u^a})={\cal W}^a_+(u^a,\overline{u^a})-{\cal W}^a_-(u^a,\overline{u^a})=\cr
&2\sgn(\alpha(\xi_E))\bigl(|C_0|^2+h\bigl(2\re(\overline{C_0}C_1)+|C_0|^2\partial_x\bigl({p_1\over\partial_xp_0}\bigr)(\xi_E)\bigr)
+{\cal O}(h^2)\bigr)\cr}$$

The condition that $u^a$ be normalized mod ${\cal O}(h^2)$ (once we have chosen $C_0$ to be real), is then
$$C_1(E)=-{1\over2}C_0\partial_x\bigl({p_1\over\partial_xp_0}\bigr)(a_E)\leqno(3.6)$$ 
so that now 
${\cal W}^a(u^a,\overline{u^a})=2\sgn(\alpha(\xi_E))C_0^2\bigl(1+{\cal O}(h^2)\bigr)$. 
We say that
$u^a$ is {\it well-normalized} mod ${\cal O}(h^2)$. This can be formalized by considering $\{a_E\}$ as a {\it Poincar\'e section}
(see Sect.4), and Poisson operator
the operator that assigns, 
in a unique way, to the initial condition $C_0$ on $\{a_E\}$ the well-normalized (forward) solution $u^a$ to $(P-E)u^a=0$: 
namely, $C_1(E)$ and $D_1(\xi)$, hence also $\widehat u^a$, depend linearly on $C_0$. 
Using the approximation 
$$C_{0}+hC_{1}(E)+hD_{1}(\xi)=\bigl(C_{0}+hC_{1}(E)+h\re(D_{1}(\xi))\bigr)\exp\bigl[{ih\over C_0}\im(D_{1}(\xi))\bigr]+{\cal O}(h^2)$$
the normalized WKB solution near $a_E$ now writes, by (3.4)
$$\widehat{u}^{a}(\xi;h)=\bigl(C_{0}+hC_{1}(E)+h\re(D_{1}(\xi))\bigr)|\alpha(\xi)|^{-{1\over2}}
\exp\bigl[i\widetilde S(\xi,\xi_E;h)/h\bigr](1+{\cal O}(h^{2}))\leqno(3.7)$$
with the $h$-dependent phase function
$$\widetilde S(\xi,\xi_E;h)=\psi(\xi)+h\int_{\xi_E}^\xi{p_{1}(-\psi'(\zeta),\zeta)\over\alpha(\zeta)}\,d\zeta+
{h^{2}\over C_{0}}\im(D_{1}(\xi))$$
The modulus of $\widehat{u}^{a}(\xi;h)$ can further be simplified using (3.6) and formula (3.10) below:
$$C_{0}+hC_{1}(E)+h\re(D_{1}(\xi))=C_0\bigl(1-{h\over2}\partial_x\bigl({p_1\over\partial_xp_0}\bigr)|_{x=-\psi'(\xi)}\bigr)=
C_0\bigl[\exp h\partial_x\bigl({p_1\over\partial_xp_0}\bigr)|_{x=-\psi'(\xi)}\bigr]^{-1/2}+{\cal O}(h^2)$$
which altogether, recalling $\alpha(\xi)=\partial_xp_0(-\psi'(\xi),\xi)$ near $\xi_E$ (and assuming $\alpha(\xi_E)>0$
to fix the ideas), gives
$$\widehat{u}^{a}(\xi;h)={1\over\sqrt2}\bigl((\partial_xp_0)\exp\bigl[h\partial_x\bigl({p_1\over\partial_xp_0}\bigr)\bigr]\bigr)^{-1/2}
\exp\bigl[i\widetilde S(\xi,\xi_E;h)/h\bigr](1+{\cal O}(h^{2}))\leqno(3.8)$$
\medskip
\noindent {\it b) The homology class of the generalized action: Fourier representation}.
\smallskip
Here we identify the various terms in (3.8), which are responsible for the holonomy of $u^a$. 
First on $\gamma_E$ (i.e. $\Lambda_E$) we have $\psi(\xi)=\int-x\, d\xi+\Const $, and $\varphi(x)=\int\xi\, dx+\Const$  
By Hamilton equations 
$$\dot\xi(t)=-\partial_xp_0(x(t),\xi(t)), \quad \dot x(t)=\partial_\xi p_0(x(t),\xi(t))$$ 
so
$\int{p_1\over\partial_x p_0}\, d\xi=-\int{p_1\over\partial_\xi p_0}\, dx=-\int_{\gamma_E} p_1\, dt$. 
The form $p_1\, dt$ is called the subprincipal 1-form.
Next we consider $D_1(\xi)$ as the integral over $\gamma_E$ of the 1-form, defined near $a$ in Fourier representation as
$$\Omega_1=T_1\,d\xi=\sgn(\alpha(\xi))\bigl(i\widetilde p_2b_0-\partial_\xi\lambda_0\bigr)|\alpha|^{-1/2} e^{-i\int{p_1\over\alpha}}\,d\xi
\leqno(3.9)$$ 
Since $\gamma_E$ is Lagrangian, $\Omega_1$ is a closed form that we are going to compute modulo exact forms. 
Using integration by parts, the integral of $\Omega_1(\xi)$ in Fourier representation simplifies to
$$\leqalignno{
&\sqrt2\re D_1(\xi)=-{1\over2}\bigl[\partial_x\bigl({p_1\over\partial_xp_0}\bigr)]_{\xi_E}^\xi
=-{1\over2}\partial_x\bigl({p_1\over\partial_xp_0}\bigr)(\xi)-{C_1(E)\over C_0}&(3.10)\cr
\sqrt2\im &D_1(\zeta)=\int_{\xi_E}^\xi T_1(\zeta)\,d\zeta+
\bigl[{\psi''\over6\alpha}\partial_x^3 p_0+{\alpha'\over4\alpha^2}\partial_x^2 p_0\bigr]_{\xi_E}^\xi&(3.11)\cr
&T_1={1\over\alpha}
\bigl(p_2-{1\over8}\partial_x^2\partial_\xi^2p_0+{\psi''\over12}\partial_x^3\partial_\xi p_0+{(\psi'')^2\over24}\bigl(\partial_x^4 p_0
\bigr)\bigr)+{1\over8}{(\alpha')^2\over\alpha^3}
\partial_x^2 p_0+{1\over6}\psi''{\alpha'\over\alpha^2}\partial_x^3 p_0\cr
&-{p_1\over\alpha^{2}}\bigl(\partial_x p_1-{p_1\over2\alpha}\partial_x^2 p_0\bigr)&(3.12)\cr
}$$
There follows:
\medskip
\noindent {\bf Lemma 3.2}: {\it Modulo the integral of an exact form in ${\cal A}$, with $T_1$ as in (3.12) we have:}
$$\eqalign{
&\re D_1(\xi)\equiv0\cr
&\sqrt2\im D_1(\xi)\equiv\int_{\xi_E}^\xi T_1(\zeta)\,d\zeta\cr
}\leqno(3.13)$$ 
Passing from Fourier to spatial representation, 
we can carry the integration in $x$-variable between the focal points $a_E$ and $a'_E$, and in $\xi$-variable again
near $a'_E$. Since $\gamma_E$ is smoothly embedded, the microlocal solution $\widehat u^a$ extends uniquely along $\gamma_E$. 

If $f(x,\xi), g(x,\xi)$ are any smooth functions on ${\cal A}$ we set
$\Omega(x,\xi)=f(x,\xi)\, dx+g(x,\xi)\, d\xi$.  
By Stokes formula
$$\int_{\gamma_E}\Omega(x,\xi)=\int\int_{p_0\leq E}(\partial_xg-\partial_\xi f)\, dx\wedge d\xi$$
where, following [CdV], we have extended $p_0$ in the disk bounded by ${\cal A}_-$ 
so that it coincides with a harmonic oscillator in a neighborhood of a point inside, say $p_0(0,0)=0$.
Making the symplectic change of coordinates $(x,\xi)\mapsto(t,E)$ in $T^*{\bf R}$:
$$\int\int_{p_0\leq E}(\partial_xg-\partial_\xi f)\, dx\wedge d\xi=\int_0^E\int_0^{T(E')}(\partial_xg-\partial_\xi f)\, dt\wedge dE'$$
where $T(E')$ is the period of the flow of Hamilton vector field $H_{p_0}$ at energy $E'$ ($T(E')$ being a constant near (0,0)). 
Taking derivative with respect to $E$, we find
$${d\over dE}\,\int_{\gamma_{E}}\Omega(x,\xi)=\int_{0}^{T(E)}(\partial_{x} g-\partial_{\xi} f)\,dt\leqno(3.14)$$
We compute $\int_{\xi_E}^\xi T_1(\zeta)\,d\zeta$ with $T_1$ as in (3.12), and start to simplify
$J_{1}=\displaystyle \int\omega_{1}$, with $\omega_1$ the last term on the RHS of (3.12). 
Let $g_{1}(x,\xi)={p_{1}^{2}(x,\xi)\over\partial_{x} p_{0}(x,\xi)}$, by (3.14) we get
$$\eqalign{
J_{1}&={1\over2}\int_{\gamma_{E}}{\partial_{x}\,g_{1}(x,\xi)\over\partial_{x}\,p_{0}(x,\xi)}\,d\xi
=-{1\over2}\int_{0}^{T(E)}\partial_{x} g_{1}(x(t),\xi(t))\,dt=
-{1\over2}{d\over dE}\int_{\gamma_{E}}g_{1}(x,\xi)\,d\xi\cr
&=-{1\over2}{d\over dE}\int_{\gamma_{E}}{p_{1}^{2}(x,\xi)\over\partial_{x} p_{0}(x,\xi)}\,d\xi
={1\over2}{d\over dE}\int_{0}^{T(E)}p_{1}^{2}(x(t),\xi(t))\,dt\cr
}\leqno(3.15)$$
which is the contribution of $p_1$ to the second term $S_2$ of generalized action in [CdV,Thm2]. Here $T(E)$ is the period on $\gamma_E$. 
We also have
$$\int_{\xi_{E}}^{\xi}{1\over\alpha(\xi)}p_{2}(-\psi'(\xi),\xi)\,d\xi=
\int_{\gamma_{E}}{p_{2}(x,\xi)\over\partial_{x}p_{0}(x,\xi)}\,d\xi=-\int_{0}^{T(E)}p_{2}(x(t),\xi(t))\,dt\leqno(3.16)$$
To compute $T_1$ modulo exact forms we are left to simplify in (3.12) the expression
$$\eqalign{
J_{2}&=\int_{\xi_{E}}^{\xi}{1\over\alpha}\bigl(-{1\over8}{\partial^{4} p_{0}\over\partial x^{2}\,\partial \xi^{2}}+
{\psi''\over12}{\partial^{4} p_{0}\over\partial x^{3}\partial \xi}+
{(\psi'')^{2}\over24}{\partial^{4} p_{0}\over\partial x^{4}}\bigr)\,d\zeta+
{1\over8}\int_{\xi_{E}}^{\xi} {(\alpha')^{2}\over\alpha^{3}}{\partial^{2} p_{0}\over\partial x^{2}}\,d\zeta\cr
&+{1\over6}\int_{\xi_{E}}^{\xi}\,\psi''{\alpha'\over\alpha^{2}}{\partial^{3} p_{0}\over\partial x^{3}}\,d\zeta\cr
}$$
Let $g_{0}(x,\xi)={\Delta(x,\xi)\over\partial_{x} p_{0}(x,\xi)}$, where we have set according to [CdV]
$$\Delta(x,\xi)={\partial^{2} p_{0}\over\partial x^{2}}{\partial^{2} p_{0}\over\partial \xi^{2}}-
\bigl({\partial^{2} p_{0}\over\partial x\,\partial \xi}\bigr)^{2}$$
Taking second derivative of eikonal equation $p_0(-\psi'(\xi),\xi)=E$, we get
$${(\partial_{x}g_{0})(-\psi'(\xi),\xi)\over\alpha(\xi)}=
{\psi'''\over\alpha}{\partial^{3} p_{0}\over\partial x^{3}}+
2\psi''{\alpha'\over\alpha^{2}}{\partial^{3} p_{0}\over\partial x^{3}}+
{\alpha''\over\alpha^{2}}{\partial^{2} p_{0}\over\partial x^{2}}-
2{\alpha'\over\alpha^{2}}{\partial^{3} p_{0}\over\partial x^{2}\partial \xi}
+{(\alpha')^{2}\over\alpha^{3}}{\partial^{2} p_{0}\over\partial x^{2}}
$$
Integration by parts of the first and third term on the RHS gives altogether
$$\eqalign{
\int_{\xi_{E}}^{\xi}{(\partial_{x} g_{0})(-\psi'(\xi),\xi)\over\alpha(\xi)}&\,d\xi=
-3\int_{\xi_{E}}^{\xi}{1\over\alpha}{\partial^{4} p_{0}\over\partial x^{2}\,\partial \xi^{2}}\,d\zeta+
2\int_{\xi_{E}}^{\xi}{\psi''\over\alpha}{\partial^{4} p_{0}\over\partial x^{3}\,\partial \xi}\,d\zeta+
\int_{\xi_{E}}^{\xi}{(\psi'')^{2}\over\alpha}{\partial^{4} p_{0}\over\partial x^{4}}\,d\zeta\cr
&+3\int_{\xi_{E}}^{\xi}{(\alpha')^{2}\over\alpha^{3}}{\partial^{2} p_{0}\over\partial x^{2}}\,d\zeta
+4\int_{\xi_{E}}^{\xi}\psi''{\alpha'\over\alpha^{2}}{\partial^{3} p_{0}\over\partial x^{3}}\,d\zeta\cr
&+\bigl[{\psi''\over\alpha}{\partial^{3} p_{0}\over\partial x^{3}}\bigr]_{\xi(E)}^{\xi}+
\bigl[{\alpha'\over\alpha^{2}}{\partial^{2} p_{0}\over\partial x^{2}}\bigr]_{\xi_{E}}^{\xi}
+3\bigl[{1\over\alpha}{\partial^{3} p_{0}\over\partial x^{2}\partial \xi}\bigr]_{\xi_{E}}^{\xi}\cr
}$$
and modulo the integral of an exact form in ${\cal A}$
$$\eqalign{
J_{2}&\equiv {1\over24}\int_{\xi_{E}}^{\xi}{(\partial_{x} g_{0})(-\psi'(\zeta),\zeta)\over\alpha(\zeta)}\,d\zeta=
-{1\over24}\,\int_{0}^{T(E)}\partial_{x} g_{0}(x(t),\xi(t))\,dt \cr
&=-{1\over24}{d\over dE}\,\int_{\gamma_{E}}g_{0}(x,\xi)\,d\xi\cr
&=-{1\over24}{d\over dE}\,\int_{\gamma_{E}}{\Delta(x,\xi)\over\partial_{x} p_{0}(x,\xi)}\,d\xi={1\over24}{d\over dE}
\int_{0}^{T(E)}\Delta(x(t),\xi(t))\,dt\cr
}$$
Using these expressions, we recover the well known action integrals (see e.g. [CdV]):
\medskip
\noindent {\bf Proposition 3.3}: {\it Let $\Gamma\, dt$ be the restriction to $\gamma_E$ of the 1-form
$$\omega_0(x,\xi)=\bigl((\partial^2_xp_0)(\partial_\xi p_0)-(\partial_x\partial_\xi p_0)(\partial_x p_0)\bigr)\, dx+
\bigl((\partial_\xi p_0)(\partial_\xi\partial_x p_0)-(\partial^2_\xi p_0)(\partial_x p_0)\bigr)\, d\xi$$ 
We have $\re \oint_{\gamma_E}\Omega_1=0$, whereas}
$$\im \oint_{\gamma_E}\Omega_1={1\over48}\bigl({d\over dE}\bigr)^2\oint_{\gamma_E}\Gamma\, dt-\oint_{\gamma_E}p_2\, dt-
{1\over2}{d\over dE}\oint_{\gamma_E}p_1^2\, dt$$
\medskip
\noindent {\it c) Well normalized QM mod ${\cal O}(h^2)$ in the spatial representation}.
\smallskip
The next task consists in extending the solutions away from $a_E$ in the spatial representation. First we 
expand $u^a(x)=(2\pi h)^{-1/2}\int e^{ix\xi/h}\widehat{u}^{a}(\xi;h)\,d\xi=(2\pi h)^{-1/2}\int e^{i(x\xi+\psi(\xi))/h}b(\xi;h)\,d\xi$ 
near $x_E$ by stationary phase (2.4) mod ${\cal O}(h^2)$, 
selecting the 2 critical points
$\xi_\pm(x)$ near $x_E$. The phase functions take the form $\varphi_\pm(x)=x\xi_\pm(x)+\psi(\xi_\pm(x))$. 
\medskip
\noindent {\bf Lemma 3.4}: In a neighborhood of the focal point $a_E$ and for $x<x_E$, the microlocal solution 
of $(P(x,hD_x;h)-E)u(x;h)=0$ is given by (with $\pm \partial_{\xi}p_{0}(x,\xi_{\pm}(x))>0$)
$$\eqalign{
&u^{a}(x;h)={1\over\sqrt{2}}\sum_{\pm}e^{\pm i\pi/4}\bigl(\pm \partial_{\xi}p_{0}(x,\xi_{\pm}(x))\bigr)^{-1/2}\cr
&\exp\bigl[{i\over h}\bigl(\varphi_{\pm}(x)-h\,\int_{x_{E}}^{x}
{p_{1}(y,\xi_{\pm}(y))\over\partial_{\xi}p_{0}(y,\xi_{\pm}(y)}\,dy\bigr)
\bigr]\,\bigl(1+h\sqrt2\bigl(C_1+D_1(\xi_{\pm}(x))+hD_2(\xi_{\pm}(x))+{\cal O}(h^{2})\bigr)\cr
}\leqno(3.17)$$
with 
$$D_2(\xi)=-{1\over 2i}\,(\psi''(\xi))^{-1}{b''_{0}(\xi)\over b_{0}(\xi)}+
{1\over8i}(\psi''(\xi))^{-2}\bigl(\psi^{(4)}(\xi)+4\psi^{(3)}(\xi)
{b'_{0}(\xi)\over b_{0}(\xi)}\bigr)-{5\over24i}\,(\psi''(\xi))^{-3}\,
(\psi^{(3)}(\xi))^{2}\leqno(3.18)$$
The quantity $\sqrt2(C_1+D_1(\xi)))$ has been computed before; with the particular choice of $C_1=C_1(E)$ in (3.6) we have:
$$\sqrt2(C_1+D_1(\xi)))=-{1\over2}\partial_x\bigl({p_1\over\partial_xp_0}\bigr)(-\psi'(\xi),\xi)+i\sqrt2\im D_1(\xi)$$
Moreover
$$\eqalign{
&{b'_{0}(\xi)\over b_{0}(\xi)}=-{\alpha'(\xi)\over 2\,\alpha(\xi)}+{ip_{1}(-\psi'(\xi),\xi)\over \alpha(\xi)}\cr
&{b''_{0}(\xi)\over b_{0}(\xi)}=\bigl(-{\alpha'(\xi)\over 2\,\alpha(\xi)}+{i\,p_{1}(-\psi'(\xi),\xi)\over\alpha(\xi)}\big)^{2}+
{d\over d\xi}\big(-{\alpha'(\xi)\over 2\,\alpha(\xi)}+{ip_{1}(-\psi'(\xi),\xi)\over \alpha(\xi)}\bigr)
}$$ 
First, we observe that $D_2(\xi_\pm(x))$ does not contribute to the homology class of the semi-classical forms defining the action,
since it contains no integral. Thus the phase in (3.17) can be replaced, mod ${\cal O}(h^3)$ by 
$$S_\pm(x_E,x;h)=x_E\xi_E+\int_{x_E}^x\xi_\pm(y)\, dy-h\int_{x_E}^{x}{p_1(y,\xi_\rho(y))\over\partial_\xi p_0(y,\xi_\rho(y)}\,dy+\sqrt2 
h^2\im\bigl(D_1(\xi_\pm(x))\bigr)\leqno(3.19)$$
with the residue of $\sqrt2\im\bigl(D_1(\xi_\pm(x))\bigr)$, mod the integral of an exact form, computed as in Lemma 3.3. 
\medskip
\noindent {\it Proof of Proposition 3.1}. We proceed by using Proposition 1.2,
and checking directly from (3.17) that normalization relations $(u^a|F^a_+)={1\over2}$
and $(u^a|F^a_-)=-{1\over2}$
hold mod ${\cal O}(h^2)$ in the spatial representation, provided $C_1(E)$ takes the value (3.6). 
So let us compute $F^a_\pm(x)$ by stationary phase as in (3.17). In Fourier representation we have
$${i\over h}[P,\chi^a]\widehat u(\xi)=(2\pi h)^{-1}\int\int 
e^{i\bigl(-(\xi-\eta)y+\psi(\eta)\bigr)/h}c(y,{\xi+\eta\over2};h)(b_0+hb_1)(\eta)\,dy\, d\eta\leqno(3.20)$$
with Weyl symbol 
$$c(x,\xi;h)\equiv c_0(x,\xi)+hc_1(x,\xi)=\bigl(\partial_\xi p_0(x,\xi)+h\partial_\xi p_1(x,\xi)\bigr)\chi'_1(x) \ \hbox{mod}\ {\cal O}(h^2)
\leqno(3.21)$$ 
Let
$$\eqalign{
u_{x}^{\pm}&(y,\eta;h)=c({x+y\over2},\eta;h)\,
\bigl(\pm\,\partial_{\xi}p_{0}(y,\xi_{\pm}(y))\big)^{-1/2}\,\exp\bigl[-i\int_{x_{E}}^{y}
{p_{1}(z,\xi_{\pm}(z))\over\partial_{\xi}p_{0}(z,\xi_{\pm}(z))}\,dz\bigr]\times\cr
&\bigl(1+h\sqrt2\bigl(C_1+D_1(\xi_{\pm}(x)\bigr)+hD_2(\xi_{\pm}(x))+{\cal O}(h^{2})\bigr)
}$$
with leading order term $u_{x}^{(0,\pm)}(y,\eta)$. Applying stationary phase (2.3) gives
$$F^{a}_{\pm}(x;h)={1\over\sqrt{2}}\,e^{\pm i\pi/4}\,e^{{i\over h}\,
\varphi_{\pm}(x)}\,\bigl(u_{x}^{\pm}\big(x,\xi_{\pm}(x);h\big)+h\,L_1u_{x}^{(0,\pm)}(x,\xi_{\pm}(x))+{\cal O}(h^{2})\bigr)
$$
which simplifies as
$$\eqalign{
F^{a}_{\pm}&(x;h)=\pm{1\over\sqrt{2}}e^{\pm i\pi/4}\exp\bigl[{i\over h}\,\bigl(\varphi_{\pm}(x)-h\,\int_{x_{E}}^{x}
{p_{1}\big(y,\xi_{\pm}(y)\big)\over\partial_{\xi}p_{0}\bigl(y,\xi_{\pm}(y)\bigr)}\,dy\bigr)
\bigr]\big(\pm\,\partial_{\xi}p_{0}(x,\xi_{\pm}(x))\big)^{1/2}\cr
&\bigl(1+hZ(\xi_\pm(x))+h{c_{1}(x,\xi_{\pm}(x))\over c_{0}(x,\xi_{\pm}(x))}+
h{2s_{\pm}(x)\,\theta_{\pm}(x)+s'_{\pm}(x)\over 2ic_{0}(x,\xi_{\pm}(x))}\bigr)\chi'_{1}(x)\cr
}$$
mod ${\cal O}(h^{2})$, where we recall $c_0,c_1$ from (3.21). Here we have set 
$$\eqalign{
&Z(\xi_\pm(x))=\sqrt2\bigl(C_1(E)+D_1(\xi_{\pm}(x))\bigr)+D_2(\xi_{\pm}(x)\cr
&s_{\pm}(x)=({\partial^{2} p_{0}\over\partial \xi^2})(x,\xi_{\pm}(x))\,\chi'_{1}(x)=\omega_{\pm}(x)\,\chi'_{1}(x)\cr
&\theta_{\pm}(x)=-{1\over\psi''(\xi_{\pm}(x))\,\alpha(\xi_{\pm}(x))}\,\bigl(i\,p_{1}\bigl(x,\xi_{\pm}(x)\bigr)-
{\psi'''(\xi_{\pm}(x))\,\alpha(\xi_{\pm}(x))+\psi''(\xi_{\pm}(x))\,
\alpha'(\xi_{\pm}(x))\over2\,\psi''(\xi_{\pm}(x))}\bigr)\cr
}$$
and used the fact that
$$c_{0}\bigl(x,\xi_{\pm}(x)\bigr)\,\bigl(\pm\,\partial_{\xi}p_{0}(x,\xi_{\pm}(x))\bigr)^{-1/2}=
\pm\,\big(\pm\,\partial_{\xi}p_{0}(x,\xi_{\pm}(x))\bigr)^{1/2}\,\chi'_{1}(x)$$
Since $\partial_{\xi} p_{0}(x,\xi_{\pm}(x))=\psi''(\xi_{\pm}(x))\,\alpha(\xi_{\pm}(x))$ we obtain
$$\eqalign{
F&^{a}_{\pm}(x;h)=\pm{1\over\sqrt{2}}\,e^{\pm i\pi/4}\,\exp\bigl[{i\over h}\bigl(\varphi_{\pm}(x)-h\,\int_{x_{E}}^{x}
{p_{1}(y,\xi_{\pm}(y))\over\partial_{\xi}p_{0}(y,\xi_{\pm}(y))}\,dy\bigr)\,
\bigr]\big(\pm\,\partial_{\xi}p_{0}(x,\xi_{\pm}(x))\big)^{1/2}\chi'_{1}(x)\cr
&\bigl(1+h\re Z(\xi_\pm(x))+h{\partial_{\xi} p_{1}(x,\xi_{\pm}(x))\over\partial_{\xi} p_{0}(x,\xi_{\pm}(x))}-
ih{\omega_{\pm}(x)\,\theta_{\pm}(x)\over\partial_{\xi} p_{0}(x,\xi_{\pm}(x))}-{ih\over2}
{{d\over dx}\big(\omega_{\pm}(x)\,\chi'_{1}(x)\big)\over\partial_{\xi} p_{0}(x,\xi_{\pm}(x))\,\chi'_{1}(x)}+{\cal O}(h^{2})\bigr)\cr
}\leqno(3.22)$$
Taking the scalar product with $u^a_\pm$ gives in particular
$$\eqalign{
(u^{a}_{+}|&F^{a}_{+})={1\over2}\,\int_{x_{E}}^{+\infty}\chi'_{1}(x)\,dx+\cr
&{h\over2}\,\int_{x_{E}}^{+\infty}\bigl(
2\re Z(\xi_\pm(x))+
{\partial_{\xi} p_{1}(x,\xi_{+}(x))\over\psi''(\xi_{+}(x))\,\alpha(\xi_{+}(x))}+{i\omega_{+}(x)
\overline{\theta_{+}(x)}\overline\psi''(\xi_{+}(x))\,\alpha(\xi_{+}(x))}\bigr)\,\chi'_{1}(x)\,dx\cr
&+{ih\over4}\,\int_{x_{E}}^{+\infty}{1\over\psi''(\xi_{+}(x))\,\alpha(\xi_{+}(x))}
{d\over dx}\big(\omega_{+}(x)\,\chi'_{1}(x)\bigr)\,dx+{\cal O}(h^{2})\cr
&={1\over2}+{h\over2}\,K_{1}+{ih\over4}\,K_{2}+{\cal O}(h^{2})\cr
}\leqno(3.23)$$
There remains to relate $K_1$ with $K_2$. We have
$$\eqalign{
&2\re Z(\xi_\pm(x))+
{\partial_{\xi} p_{1}(x,\xi_{+}(x))\over\psi''(\xi_{+}(x))\,\alpha(\xi_{+}(x))}+{i\,\omega_{+}(x)\,
\overline{\theta_{+}(x)}\over\psi''(\xi_{+}(x))\,\alpha(\xi_{+}(x))}=\cr
&{\omega_{+}(x)\over\psi''(\xi_{+}(x))\alpha(\xi_{+}(x))}
\bigl(i\,\overline{\theta_{+}(x)}+{p_{1}(x,\xi_{+}(x))\over\psi''(\xi_{+}(x))\,\alpha(\xi_{+}(x))}\bigr)=\cr
&{i\,\omega_{+}(x)\over 2\,\bigl(\psi''(\xi_{+}(x))\bigr)^{3}\,\bigl(\alpha(\xi_{+}(x))\bigr)^{2}}\,\bigl(\psi'''(\xi_{+}(x))\,\alpha(\xi_{+}(x))+
\psi''(\xi_{+}(x))\,\alpha'(\xi_{+}(x))\bigr)\cr
}\leqno(3.24)$$
whence
$$K_{1}={i\over2}\,\int_{x_{E}}^{+\infty}{\omega_{+}(x)\over\bigl(\psi''(\xi_{+}(x))\bigr)^{3}\,\big(\alpha(\xi_{+}(x))\big)^{2}}\,
\bigl(\psi'''(\xi_{+}(x))\,\alpha(\xi_{+}(x))+
\psi''(\xi_{+}(x))\,\alpha'(\xi_{+}(x))\bigr)\,\chi'_{1}(x)\,dx$$
Here we have used that 
$$\eqalign{
&2\re Z(\xi_+(x))=-\partial_x\bigl({p_1\over\partial_xp_0}\bigr)(-\psi'(\xi),\xi)+2\re D_2(\xi_+(x))\cr
&\omega_{+}(x)=\psi'''(\xi_{+}(x))\,\alpha(\xi_{+}(x))+2
\psi''(\xi_{+}(x))\,\alpha'(\xi_{+}(x))+\bigl(\psi''(\xi_{+}(x))\bigr)^{2}{\partial^{2} p_{0}\over\partial x^{2}})(x,\xi_{+}(x))\cr
}$$
On the other hand, integrating by parts gives
$$\eqalign{
K_{2}&=\bigl[{\omega_{+}(x)\,\chi'_{1}(x)\over\psi''(\xi_{+}(x))\,\alpha(\xi_{+}(x))}\bigr]_{x_{E}}^{+\infty}-
\int_{x_{E}}^{+\infty}{d\over dx}\bigl({1\over\psi''(\xi_{+}(x))\,\alpha(\xi_{+}(x))}\bigr)\,\omega_{+}(x)\,\chi'_{1}(x)\,dx\cr
&=-\int_{x_{E}}^{+\infty}{\omega_{+}(x)\over\bigl(\psi''(\xi_{+}(x))\bigr)^{3}\,\bigl(\alpha(\xi_{+}(x))\big)^{2}}\,
\bigl(\psi'''(\xi_{+}(x))\,\alpha(\xi_{+}(x))+
\psi''(\xi_{+}(x))\,\alpha'(\xi_{+}(x))\bigr)\,\chi'_{1}(x)\,dx\cr
&=2iK_{1}\cr
}$$
This shows $(u^{a}_{+}|F^{a}_{+})={1\over2}+{\cal O}(h^{2})$, and we argue similarly for 
$(u^{a}_{-}|F^{a}_{-})$, and Proposition 3.1 is proved.
\medskip
Away from $x_E$, we use 
standard WKB theory extending (3.17), with Ansatz (which we review in the Appendix)
$$u_\pm^a(x)=a_\pm(x;h)e^{i\varphi_\pm(x)/h}\leqno(3.25)$$ 
Omitting indices $\pm$
and $a$, we find $a(x;h)=a_0(x)+ha_1(x)+\cdots$; the usual half-density is
$$a_0(x)={\widetilde C_0\over C_0}|\psi''(\xi(x))|^{-1/2}b_0(\xi(x))$$
with a new constant $\widetilde C_0\in{\bf R}$~; the next term is
$$a_1(x)=(\widetilde C_1+\widetilde D_1(x))|\beta_0(x)|^{-1/2}\exp\bigl(-i\int{p_1(x,\varphi'(x))\over\beta_0(x)}\,dx\bigr)$$
and $\widetilde D_1(x)$ a complex function with 
$$\eqalign{
&\re \widetilde D_1(x)=-{1\over2}\widetilde C_0{\beta_1(x)\over\beta_0(x)}+\Const \cr
&\im \widetilde D_1(x)={\widetilde C_0}\bigl(\int{\beta_1(x)\over\beta_0^2(x)}p_1(x,\varphi'(x))\,dx
-\int{p_2(x,\varphi'(x))\over\beta_0(x)}\,dx\bigr)\cr
}\leqno(3.26)$$
and $\beta_0(x)=\partial_\xi p_0(x,\varphi'(x))=-{\alpha(\xi(x))\over\xi'(x)}$, $\beta_1(x)=\partial_\xi p_1(x,\varphi'(x))$. 
The homology class of the 1-form defining
$\widetilde D_1(x)$ can be determined as in Lemma 3.2 and coincides of course with this of $T_1\,d\xi$ (see (3.9))
on their common chart. In particular, $\im\widetilde D_1(x)=\im D_1(\xi(x))$ (where $\xi(x)$ stands for $\xi_\pm(x)$). We stress that 
(3.17) and (3.25) are equal
mod ${\cal O}(h^2)$, though they involve different expressions.

Normalization with respect to the ``flux norm'' as above yields $\widetilde C_0=C_0=1/\sqrt2$, and $\widetilde C_1$ is determined as in
Proposition 3.1. As a result
$$u(x;h)=\bigl(2\partial_\xi p_0\exp \bigl[h\partial_x\bigl({p_1\over\partial_\xi p_0}\bigr)\bigr]\bigr)^{-{1\over2}}
\exp\bigl[iS(x_E,x;h)/h\bigr](1+{\cal O}(h^{2}))\leqno(3.27)$$
This, together with (3.8), provides a covariant representation
of microlocal solutions relative to the choice of coordinate charts, $x$ and $\xi$ being related on their intersection by
$-x=\psi'(\xi)\Longleftrightarrow \xi=\varphi'(x)$. 
\medskip
\noindent {\it d) Bohr-Sommerfeld quantization rule}.
\smallskip
Recall from (3.19) the modified phase function 
of the microlocal solutions $u^a_\pm$ mod ${\cal O}(h^2)$ from the focal point $a_E$; similarly this of the other asymptotic
solution from the other focal point $a'_E$ takes the form
$$S_\pm(x'_E,x;h)=x'_E\xi'_E+\int_{x'_E}^x\xi_\pm(y)\,-h\int_{x'_E}^{x}{p_1(y,\xi_\pm(y))\over\partial_\xi p_0(y,\xi_\pm(y)}\,dy+
h^2\int_{x'_E}^{x} T_1(\xi_\pm(y))\xi'_\pm(y)\, dy\leqno(3.28)$$
Consider now
$F^{a}_\pm(x,h)$ with asymptotics (3.22), and similarly 
$F^{a'}_\pm(x,h)$. 
The normalized microlocal solutions $u^a$ and $u^{a'}$, 
uniquely extended along $\gamma_E$, are now called $u_1$ and $u_2$. Arguing as for (3.23), but taking now into account the 
variation of the semi-classical action between $a_E$ and $a'_E$ we get
$$\eqalign{
&(u_1|F_+^{a'}-F_+^{a'})\equiv{i\over2}\bigl(e^{iA_-(x_E,x'_E;h)/h}-e^{iA_+(x_E,x'_E;h)/h}\bigr)\cr
&(u_2|F_+^{a}-F_+^{a})\equiv{i\over2}\bigl(e^{-iA_-(x_E,x'_E;h)/h}-e^{-iA_+(x_E,x'_E;h)/h}\bigr)\cr
}\leqno(3.29)$$
mod ${\cal O}(h^2)$, where the generalized actions are given by
$$\eqalign{
A_\rho&(x_E,x'_E;h)=S_\rho(x_E,x;h)-S_\rho(x'_E,x;h)=\cr
&x_E\xi_E-x'_E\xi'_E+\int_{x_E}^{x'_E}\xi_\rho(y)\, dy-h\int_{x_E}^{x'_E}
{p_1(y,\xi_\rho(y))\over\partial_\xi p_0(y,\xi_\rho(y)}\,dy
+h^2\int_{x_E}^{x'_E}T_1(\xi_\rho(y))\xi'_\rho(y)\,dy\cr
}\leqno(3.30)$$
We have 
$$\eqalign{
&\int_{x'_E}^{x_E}\bigl(\xi_+(y)-\xi_-(y)\bigr)\,dy=\oint_{\gamma_E}\xi(y)\,dy\cr
&\int_{x'_E}^{x_E}\bigl({p_1(y,\xi_+(y))\over\partial_\xi p_0(y,\xi_+(y))}-{p_1(y,\xi_-(y))\over\partial_\xi p_0(y,\xi_-(y))}\bigr)
\,dy=\int_{\gamma_E}p_1\,dt\cr
&\int_{x'_E}^{x_E}\bigl(T_1(\xi_+(y))\xi'_+(y)-T_1(\xi_-(y))\xi'_-(y)\bigr)\,dy=\im\oint_{\gamma_E}\Omega_1(\xi(y))\,dy\cr
}$$
On the other hand, Gram matrix as in (2.7)
has determinant 
$$-\cos^2 \bigl((A_-(x_E,x'_E;h)-A_+(x_E,x'_E;h))/2h)$$ 
which vanishes precisely when BS holds. This brings our alternative proof of Theorem 0.1 to an end. 

\bigskip
\noindent {\bf 4. Bohr-Sommerfeld and action-angle variables}.
\smallskip
We present here a simpler approach based on Birkhoff normal form and the monodromy operator
[LoRo], which reminds of [HeRo].
Let $P$ be self-adjoint as in (0.1) with Weyl symbol $p\in S^0(m)$, and such that there exists a topological ring ${\cal A}$
where $p_0$ verifies the hypothesis (H) in the Introduction. Without loss of generality,
we can assume that $p_0$ has a periodic orbit $\gamma_0\subset{\cal A}$ with period $2\pi$ and energy $E=E_0$. 
Recall from Hamilton-Jacobi theory that there exists a smooth canonical transformation 
$(t,\tau)\mapsto\kappa(t,\tau)=(x,\xi)$, $t\in[0,2\pi]$,
defined in a neighborhood of $\gamma_0$ and a smooth function 
$\tau\mapsto f_0(\tau)$, $f_0(0)=0$, $f'_0(0)=1$ such that 
$$p_0\circ\kappa(t,\tau)=f_0(\tau)\leqno(4.1)$$
It is given by its generating function $S(\tau,x)=\int_{x_0}^x\xi\,dx$, $\xi=\partial_xS$, $\varphi=\partial_\tau S$,
and
$$p_0(x,{\partial S\over\partial x}(\tau,x))=f_0(\tau)\leqno(4.2)$$
Energy $E$ and momentum $\tau$ are related by the 1-to-1 transformation $E=f_0(\tau)$, and $f'_0(E_0)=1$.

This map can be quantized semi-classically, which is known as the semi-classical Birkhoff normal form (BNF), see e.g. [GuPa]
and its proof. Here we take advantage of the fact (see [CdV], Prop.2) that we can deform smoothly $p$ in the interior of annulus ${\cal A}$,
without changing its semi-classical spectrum in $I$, in such a way that the ``new'' $p_0$ has a non-degenerate
minimum, say at $(x_0,\xi_0)=0$, with $p_0(0,0)=0$, while all energies $E\in]0,E_+]$ are regular.
Then BNF can be achieved by introducing
the so-called ``harmonic oscillator'' coordinates $(y,\eta)$ so that (4.1) takes the form
$$p_0\circ\kappa(y,\eta)=f_0({1\over2}(\eta^2+y^2))\leqno(4.3)$$
and $U^*PU=f({1\over2}\bigl((hD_y)^2+y^2\bigr);h)$,
has full Weyl symbol
$f(\tau;h)=f_0(\tau)+hf_1(\tau)+\cdots$.
Here $f_1$ includes Maslov correction 1/2,
and $U$ is a microlocally unitary $h$-FIO operator associated with $\kappa$ ([CdVV], [HeSj]). 
In ${\cal A}$, $\tau\neq0$, so we can make the smooth symplectic change of coordinates $y=\sqrt{2\tau}\cos t$, 
$\eta=\sqrt{2\tau}\sin t$,
and take ${1\over2}\bigl((hD_y)^2+y^2\bigr)$ back to $hD_t$. 

We do not intend to provide an explicit expression for $f_j(\tau)$, $j\geq1$ in term of the $p_j$,
but only point out that $f_j$ depends linearly on $p_0,p_1,\cdots p_j$ and their derivatives. Of course,
BNF allows to get rid of focal points. 
The section $t=0$ in $f_0^{-1}(E)$ (Poincar\'e section) reduces to a point, say $\Sigma=\{a(E)\}$. 

Recall from [LoRo] that Poisson operator ${\cal K}(t,E)$ here solves (globally near $\gamma_0$)
$$(f(hD_t;h)-E){\cal K}(t,E)=0\leqno(4.4)$$
and is given in the special 1-D case by the multiplication operator on $L^2(\Sigma)\approx{\bf C}$
$${\cal K}(t,E)=e^{iS(t;E)/h}a(t;E,h)$$ 
where 
$S(t,E)$ verifies the eikonal equation
$f_0(\partial_tS)=E$, $S(0,E)=0$, i.e. $S(t,E)=f_0^{-1}(E)t$, and $a(t,E;h)=a_0(t,E)+ha_1(t,E)+\cdots$ satisfies transport equations 
to any order in $h$. 

Applying (3.25) in the special case where $P$ has constant coefficients, one has
$$\eqalign{
&a_0(t,E)=C_0\bigl((f_0^{-1})'(E)\bigr)^{1/2}e^{it\widetilde S_1(E)}\cr
&a_1(t,E)=\bigl(C_1(E)+C_0(\beta(E)+it\widetilde S_2(E))\bigr)\bigl((f_0^{-1})'(E)\bigr)^{1/2}e^{it\widetilde S_1(E)}\cr
}\leqno(4.5)$$
with $C_0\in{\bf R}$ a normalization constant as above to be determined as above
$$\eqalign{
&\widetilde S_1(E)=-f_1(\tau)(f_0^{-1})'(E)\cr
&\beta(E)=-{1\over2}(f_0^{-1})'(E)f'_1(\tau)\cr
&\widetilde S_2(E)=(f_0^{-1})'(E)\bigl({1\over2}{df_1^2\over dE}-f_2(\tau)\bigr)\cr
}\leqno(4.6)$$
where we recall $\tau=f_0^{-1}(E)$, so that
$${\cal K}(t,E)=e^{iS(t;E)/h}\bigl((f_0^{-1})'(E)\bigr)^{1/2}e^{it\widetilde S_1(E)}\bigl(C_0
+hC_1(E)+hC_0\beta(E)+ithC_0\widetilde S_2(E)\bigr)\leqno(4.7)$$ 
Together with ${\cal K}(t,E)$ we define ${\cal K}^*(t,E)=e^{-iS(t,E)/h}\overline{a(t,E;h)}$, and 
$${\cal K}^*(E)=\int {\cal K}^*(t,E)\, dt$$ 
The ``flux norm'' on ${\bf C}^2$ is defined by
$$(u|v)_\chi=\bigl({i\over h}[f(hD_t;h),\chi(t)]{\cal K}(t;h)u|{\cal K}(t,h)v\bigr)\leqno(4.8)$$ 
with the scalar product of $L^2({\bf R}_t)$ on the RHS, and $\chi\in C^\infty({\bf R})$ is a smooth step-function, 
equal to 0 for $t\leq0$ and to 1 for $t\geq2\pi$. To normalize ${\cal K}(t,E)$ we start from
$${\cal K}^*(E){i\over h}[f(hD_t;h),\chi(t)]{\cal K}(t,E)=\Id_{L^2({\bf R})}$$
Since ${i\over h}[f(hD_t;h),\chi(t)]$ has Weyl symbol $(f'_0(\tau))+hf'_1(\tau))\chi'(t)+{\cal O}(h^2)$ we are led
to compute 
$I(t,E)={i\over h}[f(hD_t;h),\chi(t)]{\cal K}(t,E)$ where we have set $Q(\tau;h)=f'_0(\tau)+hf'_1(\tau)$. 
Again by stationary phase (2.3)
$$\eqalign{
I(t,E)&=e^{iS(t,E)/h}\bigl[Q(\tau;h))\chi'(t)a(t,E;h)-ih\partial_\tau Q(\tau;h)\bigl({1\over2}\chi''(t)a(t,E;h)\cr
&+\chi'(t)\partial_ta(t,E;h)+{\cal O}(h^2)\bigr]\cr
}$$
Integrating $I(t,E)$ against $e^{-iS(t,E)/h}\overline{a(t,E;h)}$, we get
$$\eqalign{
(u|v)_\chi&=u\overline v\bigl[Q(\tau;h)\int\chi'(t)|a(t,E;h)|^2-{ih\over2}\partial_\tau Q(\tau;h)
\int\chi''(t)|a(t,E;h)|^2\,dt\cr
&-ih\partial_\tau Q(\tau;h)
\int\partial_ta(t,E;h)\overline{a(t,E;h)}\chi'(t)\,dt
+{\cal O}(h^2)\bigr]
}\leqno(4.9)$$
Now $|a(t,E;h)|^2=(f_0^{-1})'(E)\bigl(C_0^2+2hC_0C_1(E)+2hC_0^2\beta(E)\bigr)+{\cal O}(h^2)$ is independent of $t$
mod ${\cal O}(h^2)$, and 
$$(u|v)_\chi=u\overline v\bigl(C_0^2+2C_0C_1(E)h-C_0^2\alpha(E)(f_0^{-1})'(E)f''_0(\tau)+{\cal O}(h^2)\bigr)$$
so that, choosing $C_0=1$ and
$$C_1(E)={1\over2}\bigl((f_0^{-1})'(E)\bigr)^2f_1(\tau)f''_0(\tau)$$
we end up with $(u|v)_\chi=u\overline v(1+{\cal O}(h^2)$, which normalizes ${\cal K}(t,E)$ to order 2.
 
We define ${\cal K}_0(t,E)={\cal K}(t,E)$
(Poisson operator with data at $t=0$), ${\cal K}_{2\pi}(t,E)={\cal K}(t-2\pi,E)$ (Poisson operator with data at $t=2\pi$),
and recall from [LoRo] that $E$ is an eigenvalue of $f(hD_t;h)$ iff 1 is an eigenvalue of the monodromy operator
$M(E)=K_{2\pi}^*(E){i\over h}[f(hD_t;h),\chi]K_0(\cdot,E)$, which in the 1-D case reduces again to a multiplication operator. 
A short computation shows that
$$M(E)=\exp[2i\pi\tau/h]\exp[2i\pi\widetilde S_1(E)]\bigl(1+2i\pi h\widetilde S_2(E)+{\cal O}(h^2)\bigr)$$
so again BS quantization rule writes with an $h^2$ accuracy as
$$f_0^{-1}(E)+h\widetilde S_1(E)+h^2\widetilde S_2(E)\equiv nh, \ n\in{\bf Z}$$
Let $S_1(E)=2\pi\widetilde S_1(E)$, and $S_2(E)=2\pi\widetilde S_2(E)$. 
Since $f_0^{-1}(E)=\tau(E)={1\over2\pi}\oint_{\gamma_E}\xi\,dx$, and we know that $S_3(E)=0$,
we eventuelly get 
$$S_0(E)+hS_1(E)+h^2S_2(E)+{\cal O}(h^4)=2\pi n h, \ n\in{\bf Z}$$

Note that the proof above readily extends to the periodic case, where there is no Maslov correction in $f_1$. 


\bigskip
\noindent {\bf 5. The discrete spectrum of $P$ in $I$}.
\smallskip
Here we recover the fact that BS determines asymptotically all eigenvalues of $P$ in $I$. As in Sect.1 we adapt the argument of [SjZw], 
and content ourselves with the computations below with an accuracy ${\cal O}(h)$. 
It is convenient to think of $\{a_E\}$ and $\{a'_E\}$ as 
zero-dimensional ``Poincar\'e sections'' of 
$\gamma_E$.
Let ${\cal K}^a(E)$ be the operator (Poisson operator) that assigns to its ``initial value'' 
$C_0\in L^2(\{a_E\})\approx{\bf R}$ the well
normalized solution $u(x;h)=\int e^{i(x\xi+\psi(\xi))/h}b(\xi;h)\, d\xi$
to $(P-E)u=0$ near $\{a_E\}$. By construction, we have:
$$\pm {\cal K}^a(E)^*{i\over h}[P,\chi^a]_\pm {\cal K}^a(E)=\Id_{a_E}=1\leqno(5.1)$$
We define objects ``connecting'' $a$ to $a'$ along $\gamma_E$ as follows: let $\widetilde T=\widetilde T(E)>0$ such that 
$\exp \widetilde TH_{p_0}(a)=a'$ (in case $p_0$ is invariant by time reversal, i.e.
$p_0(x,\xi)=p_0(x,-\xi)$ we take $\widetilde T(E)=T(E)/2$).
Choose $\chi^a_f$ ($f$ for ``forward'') be
a cut-off function supported microlocally near $\gamma_E$, equal to 0 along
$\exp tH_{p_0}(a)$ for $t\leq \e $, equal to 1 along $\gamma_E$ for
$t\in[2\e ,\widetilde T+\e ]$, and back to 0 next to $a'$, e.g. for $t\geq \widetilde T+2\e $. 
Let similarly $\chi^a_b$ ($b$ for ``backward'') 
be a cut-off function supported microlocally near $\gamma_E$, equal to 1 along 
$\exp tH_{p_0}(a)$ for $t\in[-\e ,\widetilde T-2\e ]$, and equal to 0 next to $a'$, e.g. for $t\geq \widetilde T-\e $. 
By (5.1) we have 
$$\leqalignno{
&{\cal K}^a(E)^*{i\over h}[P,\chi^a]_+{\cal K}^a(E)={\cal K}^a(E)^*{i\over h}[P,\chi^a_f]{\cal K}^a(E)=1&(5.2)\cr 
&-{\cal K}^a(E)^*{i\over h}[P,\chi^a]_-{\cal K}^a(E)=-{\cal K}^a(E)^*{i\over h}[P,\chi^a_b]{\cal K}^a(E)=1&(5.3)\cr
}$$ 
which define a left inverse $R_+^a(E)={\cal K}^a(E)^*{i\over h}[P,\chi^a_f]$ to ${\cal K}^a(E)$ and a right inverse 
$$R_-^a(E)=-{i\over h}[P,\chi^a_b]{\cal K}^a(E)$$ 
to ${\cal K}^a(E)^*$.
We define similar objects connecting $a'$ to $a$, $\widetilde T'=\widetilde T'(E)>0$ such that $\exp \widetilde T'H_{p_0}(a)=a'$ ($\widetilde T=
\widetilde T'$
if $p_0$ is invariant by time reversal), in particular a
left inverse $R_+^{a'}(E)={\cal K}^{a'}(E)^*{i\over h}[P,\chi^{a'}_f]_+$ to ${\cal K}^{a'}(E)$ and a right inverse 
$R_-^{a'}(E)=-{i\over h}[P,\chi^{a'}_b]{\cal K}^{a'}(E)$ to ${\cal K}^{a'}(E)^*$, with the additional requirement 
$$\chi^a_b+\chi^{a'}_b=1\leqno(5.4)$$ 
near $\gamma_E$. 
Define now the pair $R_+(E)u=(R^a_+(E)u,R^{a'}_+(E)u)$, $u\in L^2({\bf R})$ and $R_-(E)$ by $R_-(E)u_-=R^a_-(E)u^a_-+R^{a'}_-(E)u^{a'}_-$, 
$u_-=(u^{a}_-,u^{a'}_-)\in{\bf C}^2$, we call Grushin operator ${\cal P}(z)$ the operator defined by the linear system
$${i\over h}(P-z)u+R_-(z)u_-=v, \quad R_+(z)u=v_+\leqno(5.5)$$
From [SjZw], we know that the problem (5.5) is well posed, and as in (1.7)-(1.8)
$${\cal P}(z)^{-1}=\pmatrix{E(z)&E_+(z)\cr E_-(z)&E_{-+}(z)}$$ 
with choices of $E(z),E_+(z),E_{-+}(z),E_-(z)$ similar to those in Sect.1. Actually one can show that the effective Hamiltonian $E_{-+}(z)$ 
is singular precisely when 1 belongs to the spectrum of the monodromy operator, or
when the microlocal solutions $u_1,u_2\in K_h(E)$ computed in (3.29) are colinear, which amounts to say that 
Gram matrix (2.7) is singular.
There follows that the spectrum of $P$ in $I$ is precisely the set of $z$ we have determined by BS quantization rule.

Note that the argument used in Sect.4 would need a slightly different justification, since we made use of a single
``Poincar\'e section''. 

\bigskip
\noindent {\bf Appendix: Essentials on 1-D semi-classical spectral asymptotics}.
\smallskip
Following essentially [BaWe] [CdV2], we recall here some useful notions of 1-D Microlocal Analysis,
providing a consistent framework for WKB expansions in different representations. 
\medskip
\noindent {\it a) $h$-Pseudo-differential Calculus}
\smallskip
Semi-classical analysis, or $h$-Pseudodifferential calculus, 
is based on asymptotics with respect to the small parameter $h$. This
is a (almost straightforward) generalization of the Pseudo-differential calculus of [H\"o], based on 
asymptotics with respect to smoothness, that we refer henceforth as the ``Standard Calculus''. 

The growth at infinity of an Hamiltonian is controlled by an 
order function, i.e. $m\in C^\infty(T^*{\bf R})$, $m\geq1$, of temperate growth at infinity, that verifies $m\in S(m)$; for instance 
we take $m(x,\xi)= 1+|\xi|^2$
for Schr\"odinger or Helmholtz Hamiltonians with long range potential,
$m(x,\xi)= 1+|x,\xi|^2$ for Hamiltonians of the type of a harmonic oscillator (with compact resolvant), or simply $m=1$
for a phase-space ``cut-off''. 
    
Consider a real valued symbol $p\in S(m)$ as in (0.1), and define a self-adjoint 
$h$-PDO $p^w(x,hD_x;h)$ on $L^2({\bf R})$ as in (0.3). 

As in the Standard Calculus, $h$-PDO's compose in a natural way. 
It is convenient to work with symbols having asymptotic expansions (0.2).
A $h$-PDO $P^w(x,hD_x;h)$ 
is called elliptic 
if its principal symbol $p_0$ verifies $|p_0(x,\xi)|\geq \const m(x,\xi)$. If $P^w(x,hD_x;h)$ is elliptic then it has an 
inverse $Q^w(x,hD_x;h)$ with $q\in S(1/m)$. Ellipticity can be restricted in the microlocal sense,
i.e. we say that $p$ is elliptic at $\rho_0=(x_0,\xi_0)\in T^*{\bf R}$ if $p_0(\rho_0)\neq0$, so that $P^w(x,hD_x;h)$ has 
also a microlocal inverse 
$Q^w(x,hD_x;h)$ near $\rho_0$. 
\medskip
\noindent {\it b) Admissible semi-classical distributions and microlocalization}
\smallskip
These $h$-PDO extend naturally by acting 
on spaces of distributions of finite regularity $H^s({\bf R})$ (Sobolev spaces). 

It is convenient to view $h$-PDO's as acting on a family $(u_{h})$ of $L^2$-functions, or distributions on ${\bf R}$, rather
than on individual functions. We call $u_h$ {\it admissible} iff for any compact set $K\subset{\bf R}$ we have $\|u_h\|_{H^s(K)}={\cal O}(h^{-N_0})$
for some $s$ and $N_0$. We shall be working with some particular admissible distributions, called {\it Lagrangian distributions},
or oscillating integrals. 

A Lagrangian distribution takes the form
$$u_h(x)=(2\pi h)^{-N/2}\int_{{\bf R}^N}e^{i\varphi(x,\theta)/h}a(x,\theta;h)\,d\theta\leqno(A.1)$$
where $a$ is a symbol (i.e. belongs to some $S(m)$) and $\varphi$ is a non-degenerate phase function, i.e. 
$d_{x,\theta}\varphi(x_0,\theta_0)\neq0$, 
and $d\partial_{\theta_1}\varphi,\cdots,d\partial_{\theta_N}\varphi$ are linearly independent on 
the critical set 
$$C_\varphi=\{(x,\theta):{\partial \varphi\over\partial\theta}(x,\theta)=0\}\leqno(A.2)$$  
Such a distribution is said to be {\it negligible} iff for any compact set $K\subset{\bf R}$, and any $s\in{\bf R}$ we have
$\|u_h\|_{H^s(K)}={\cal O}(h^\infty)$.
\smallskip
\noindent {\it Remark}: 
Negligible Lagrangian distributions up to finite order, as those constructed in this paper, can be defined similarly. 
Including more general admissible distributions requires to modify the concept of negligible distributions, as well as the frequency set below, 
in order to take additional regularity into
account. The way to do it is to compactify the usual phase-space $T^*{\bf R}$ by ``adding a sphere'' at infinity [CdV2]. 
For simplicity, we shall be content with microlocalizing in $T^*{\bf R}$, let us only mention that microlocalization
in case of Standard Calculus is carried in $T^*{\bf R}\setminus0$, where the zero-section has been removed, and the phase
functions enjoy certain homogeneity properties in the phase variables.
\medskip
Microlocal Analysis specifies further the ``directions'' in $T^*{\bf R}$ where $u_h$ is ``negligible''. To this end,
we introduce, following Guillemin and Sternberg, the {\it frequency set} 
$\FS u_h\subset T^*{\bf R}$ by saying that $\rho_0=(x_0,\xi_0)\notin FS u_h$ iff there exists
a $h$-PDO $A$ with symbol $a\in S^0(m)$ elliptic at $\rho_0$ and such that $Au_h$ is negligible. 
Since this definition doesn't depend of the choice of $A$,
and we can take $A=\chi^w(x,hD_x)$ where $\chi\in C^\infty_0(T^*{\bf R})$ is a microlocal cut-off equal to 1 near $\rho_0$. 
On the set of admissible distributions, we define an equivalence relation at $(x_0,\xi_0)\in T^*{\bf R}$
by $u_h\sim v_h$ iff $(x_0,\xi_0)\notin\FS (u_h-v_h)$, and we say that
$u_h=v_h$ microlocally near $(x_0,\xi_0)$. 

As in Standard Calculus, if $P\in S(m)$ we have 
$$\FS Pu_h\subset \FS u_h\subset \FS Pu_h\cup \char P\leqno(A.3)$$
where $\char P=\{(x,\xi)\in T^*{\bf R}: p_0(x,\xi)=0\}$ is the bicharacteristic strip. 

For instance, eigenfunctions of $P^w(x,hD_x;h)$ with energy $E$ (as admissible distributions) 
or more generally, solutions, in the microlocal sense, of $(P^w(x,hD_x;h)-E)u_h\sim0$
are ``concentrated'' microlocally in the energy shell $p_0(x,\xi)=E$, in the sense that
$\FS u_h\subset\char (P-E)$. It follows that $\FS u_h$ is invariant under the flow $t\mapsto \Phi^t$ of Hamilton vector field $H_{p_0}$.
Assume now that $P-E$ is of principal type (i.e. $H_{p_0}\neq0$ on $p_0=E$), the microlocal kernel of $P-E$ is (at most) one-dimensional,
i.e. if $u_h,v_h$ are microlocal solutions and $u_h\sim v_h$ at one point $(x_0,\xi_0)$, then $u_h\sim v_h$ everywhere. 
The existence of WKB solutions (see below) ensures that the microlocal kernel of $P-E$ is indeed one-dimensional.
This fails of course to be true in case of multiple caracteristics, e.g. at a separatrix. 

It is convenient to characterize the frequency set in terms of $h$-Fourier transform 
$${\cal F}_hu_h(\xi)=(2\pi h)^{-1/2}\int e^{-ix\xi/h}u_h(x)\,dx\leqno(A.4)$$
Namely $\rho_{0}\notin \FS _h(u_{h})$ iff 
there exists $\chi\in C^{\infty}_{0}({\bf R})$, $\chi(x_{0})\neq0$, and a compact neighborhood $V$ of $\xi_{0}$ such that
${\cal F}_{h}(\chi\,u_{h})(\xi)={\cal O}(h^{\infty})$ uniformly on $V$.

Note as above that the frequency set may include the zero section $\xi=0$, contrary to the standard wave-front WF, see also [Iv]. 
\smallskip
\noindent{\it Examples}:
\smallskip
1) ``WKB functions'' of the form  $u_{h}(x)=a(x)\,e^{iS(x)/h}$ with $a, S \in C^{\infty}$, $S$ real valued. We have
$\FS _h(u_{h})=\big\{(x,S'(x)): x\in\supp (a)\big\}$. More generally, if $u_h$ is as in (A.1) then
$\FS _h(u_h)$ is contained in the Lagrangian manifold $\Lambda_\varphi=\{(x,\partial_x\varphi(x,\theta)): \partial_\theta\varphi(x,\theta)=0\}$,
with equality if $a(x,\theta;h)\neq0$ on the critical set $C_\varphi$ was defined in (A.2). 

2) If $u(x)$ is independent of $h$, then
$\FS _h(u)=\WF u\cup (\supp (u)\times \{0\})$. 
\medskip
Fourier inversion formula then shows that if 
$U\subset {\bf R}^{n}$ is an open set, an $h$-admissible family $(u_{h})$ is negligible
in $U$ iff $\pi_x(\FS(u_{h}))\cap U=\emptyset$, where $\pi_x$ denotes the projection $T^*{\bf R}\to{\bf R}_x$. 
So $\FS u_h=\emptyset$ iff $u_h$ are smooth and small (with respect to $h$) in Sobolev norm. 
\medskip
\noindent {\it c) WKB method}
\smallskip
When $P-E$ is of principal type, and $H_{p_0}$ is transverse to the fiber in $T^*{\bf R}$, we seek for microlocal solutions 
of WKB type, of the form $u_h(x)=e^{iS(x)/h}a(x;h)$, where $a(x;h)\sim\Sum_{j=0}^\infty{h\over i}^ja_j(x)$. Applying
$P-E$, we get an asymptotic sum, with leading term $p_0(x,S'(x))=E$, which is the eikonal equation, that we solve
by prescribing the initial condition $S'(x_0)=\xi_0$, where $p_0(x_0,\xi_0)=E$. The lower order terms are given by
(in-)homogeneous transport equations, the first transport equation takes the invariant form ${\cal L}_{H_{p_0}}a_0=0$,
where ${\cal L}_{H_{p_0}}$ denote Lie derivative along $H_{p_0}$. Hence $e^{iS(x)/h}a_0(x)$ gives the Lagrangian manifold 
$\Lambda_S$ together with the half density $a_0(x)\sqrt{dx}$ on it. The right hand side of
higher order (non-homegeneous) transport equations or order $j$
involve combinations of previous $a_0,\cdots,a_{j-1}$.

When $H_{p_0}$ turns vertical, we switch to Fourier representation as in Sect.3. Matching of solutions in such different charts
can be done using Gram matrix since, $P-E$ being of principal type, there is only one degree of freedom for choosing the microlocal solution. 
\bigskip

\noindent {\bf References}
\smallskip
\noindent [Ar] P.Argyres. The Bohr-Sommerfeld quantization rule and Weyl correspondence, Physics 2, p.131-199 (1965)

\noindent [B] H.Baklouti. Asymptotique des largeurs de resonances pour un mod\`ele d'effet tunnel microlocal. Ann. Inst. H.Poincar\'e (Phys.Th.)
68 (2), p.179-228, 1998.

\noindent [BaWe] S.Bates, A.Weinstein. Lectures on the geometry of quantization. Berkeley Math. Lect. Notes 88,
American Math. Soc. 1997

\noindent [BenOrz] C.Bender S.Orzsag. Advanced Mathematical Methods for Scientists and Engineers. Srpinger, 1979.

\noindent [BenIfaRo] A.Bensouissi, A.Ifa, M.Rouleux. Andreev reflection and the
semi-classical Bogoliubov-de Gennes Hamiltonian.
Proceedings ``Days of Diffraction 2009'', Saint-Petersburg. p.37-42. IEEE 2009.

\noindent [BenMhaRo] A.Bensouissi, N.M'hadbi, M.Rouleux. Andreev reflection and the
semi-classical Bogoliu- bov-de Gennes Hamiltonian: resonant states.
Proceedings ``Days of Diffraction 2011'', Saint-Peters- burg. p.39-44. IEEE 101109/ DD.2011.6094362 

\noindent [BoFuRaZe] J.-F. Bony, S.Fujiie, T.Ramond, M.Zerzeri. {\bf 1}. Quantum monodromy for a homoclinic orbit. Proc.
Colloque EDP Hammamet, 2003. {\bf 2}. Resonances for homoclinic trapped sets,  arXiv:1603.07517v1.

\noindent [CaGra-SazLittlReiRios] M.Cargo, A.Gracia-Saz, R.Littlejohn, M.Reinsch \& P.de Rios. 
Moyal star product approach to the Bohr-Sommerfeld approximation, J.Phys.A: Math and Gen.38, p.1977-2004 (2005).

\noindent [Ch] J. Chazarain. Spectre d'un
Hamiltonien quantique et Mecanique classique. Comm. Part. Diff. Eq. 6, p.595-644 (1980)

\noindent [CdV] Y.Colin de Verdi\`ere. {\bf 1}. Bohr Sommerfeld rules to all orders. Ann. H.Poincar\'e, 6, p.925-936, 2005.
{\bf 2}. M\'ethodes semi-classiques et th\'eorie spectrale. 
https://www-fourier.ujf-grenoble.fr/~ycolver/ All-Articles/93b.pdf

\noindent [CdVV] Y.Colin de Verdi\`ere, J.Vey. Le lemme de Morse isochore. Topology, 18, p.283-293, 1979.

\noindent [CdVPa] Y.Colin de Verdi\`ere, B.Parisse. 

\noindent [DePh] E.Delabaere, F.Pham. Resurgence methods in semi-classical asymptotics. Ann. Inst. H.Poin- car\'e 71(1), p.1-94, 1999.

\noindent [DeDiPh] E.Delabaere, H.Dillinger, F.Pham. Exact semi-classical expansions for 1-D quantum oscillators. J.Math.Phys. Vol.38 (12)
p.6126-6184 (1997)

\noindent [Dui] J.J.Duistermaat. Oscillatory integrals, Lagrange immersions and unfolding of singularities. C.P.A.M. 27, p.207-281,
1982.

\noindent [DuGy] K.Duncan, B.Gy\"orffy. Semiclassical theory of Quasiparticlesin the superconduting state. Ann. Phys. 298,p.273-333, 2002.

\noindent [FaLoRo] H.Fadhlaoui, H.Louati, M.Rouleux. Hyperbolic Hamiltonian flows and the semiclassical Poincar\'e map. 
Proceedings ``Days of Diffraction 2013'', Saint-Petersburg, IEEE 10.1109/ DD.2013. 6712803, p.53-58.

\noindent [Fe] M.V.Fedoriouk. M\'ethodes asymptotiques pour les Equations Diff\'erentielles Ordinaires. Ed. MIR, Moscou, 1987.
(=Asymptotic Analysis. Springer, 1993.)

\noindent [Gra-Saz] A.Gracia-Saz. The symbol of a function of a pseudo-differential operator. Ann. Inst. Fourier, 55(7), p.2257-2284, 2005.

\noindent [GuPa] V.Guillemin, T.Paul, Some remarks about semiclassical trace invariants and quantum normal forms. 
Comm. Math. Phys. 294 (2010), no. 1, 1–19.

\noindent [HeRo] B.Helffer, D.Robert. Puits de potentiel generalis\'{e}s et asymptotique semi-classique. Annales
Inst. H.Poincar\'{e} (Physique Th\'{e}orique), Vol.41, No 3, p.291-331, 1984.

\noindent [HeSj] B.Helffer, J.Sj\"ostrand. Semi-classical analysis for Harper's equation III. Memoire  No 39, Soc. Math. de France, 117 (4), 
1988.

\noindent [H\"o] L.H\"ormander, The Analysis of Partial Differential Operators, I. Springer, 1983.

\noindent [IfaM'haRo] A.Ifa, N.M'hadbi, M.Rouleux. On generalized Bohr-Sommerfeld quantization rules for operators with PT symmetry. 
Math. Notes. Vol.99, No.5, p.676-684, 2016.

\noindent [IfaRo] A.Ifa, M.Rouleux. Regular Bohr-Sommerfeld quantization rules for a $h$-pseudo-differential
operator: the method of positive commutators. 
Int. Conference Euro-Maghreb Laboratory of Math. and their Interfaces, Hammamet (Tunisie). ARIMA,
Vol.23. Special issue LEM21-2016.

\noindent [Iv] V.Ivrii. Microlocal Analysis and Precise Spectral Asymptotics. Springer-Verlag, Berlin, 1998.

\noindent [Li] R.Littlejohn, Lie Algebraic Approach to Higher-Order Terms, Preprint June 2003.

\noindent [LoRo] H. Louati, M.Rouleux. {\bf 1}. Semi-classical resonances associated with a periodic orbit. 
Math. Notes, Vol. 100, No.5,  p.724-730, 2016. {\bf 2}. Semi-classical quantization rules for a periodic orbit of hyperbolic type. 
Proceedings ``Days of Diffraction 2016'', Saint-Petersburg, p.112-117, 
DOI: 10.1109/DD.2016.7756858, IEEE.

\noindent [NoSjZw] S.Nonnenmacher, J.Sj\"ostrand, M.Zworski. From Open Quantum Systems to Open Quantum maps.
Comm. Math. Phys. 304, p.1-48, 2011

\noindent [Ol] F.Olver. Asymptotics and special functions. Academic Press, 1974.

\noindent [Ro] M.Rouleux. Tunneling effects for $h$-Pseudodifferential Operators, Feshbach Resonances and the 
Born-Oppenheimer Approximation {\it in}: Evolution Equations, Feshbach
Resonances, Singular Hodge Theory. Adv. Part. Diff. Eq. Wiley-VCH (1999)

\noindent [Sj] J.Sj\"ostrand. {\bf 1}. Analytic singularities of solutions of boundary value problems. Proc. NATO ASI on 
Singularities in boundary value problems, D.Reidel, 1980, p.235-269.
{\bf 2}. Density of states oscillations for magnetic Schr\"odinger operators, 
{\it in}: Bennewitz (ed.) Diff. Eq. Math. Phys. 1990. Univ. Alabama, Birmingham, p.295-345. 

\noindent [SjZw] J. Sj\"ostrand and M. Zworski. Quantum monodromy and semi-classical trace formulae, J. Math. Pure Appl.
81, p.1-33, 2002. Erratum: http://math.berkeley.edu/~zworski/qmr.pdf

\noindent [Vo] A.Voros. Asymptotic h-expansions of stationary quantum states, Ann. Inst. H. Poincar\'{e} Sect. A(N.S), 26, p.343-403, 1977.

\noindent [Ya] D.Yafaev. The semi-classical limit of eigenfunctions of the Schr\"odinger equation and the Bohr-Sommerfeld
quantization condition, revisited. St. Petersburg Math. J., 22:6, p.1051- 1067, 2011.


\bye